\documentclass[acmsmall,screen]{acmart} % \documentclass[acmsmall,anonymous,screen, review]{acmart} % \documentclass[acmsmall,manuscript,anonymous,screen, review]{acmart} % [from CHI 2021] 
\usepackage{booktabs}
\usepackage{graphicx}
\usepackage{listings}
\usepackage{threeparttable}
\usepackage{xcolor}
\usepackage{array}
\usepackage{ragged2e}
\usepackage{hhline}
\usepackage{colortbl}
\usepackage{svg}
\usepackage{amsmath}
\usepackage{hyperref}
\usepackage{microtype}
\AtBeginDocument{%
  \providecommand\BibTeX{{%
    \normalfont B\kern-0.5em{\scshape i\kern-0.25em b}\kern-0.8em\TeX}}}

\usepackage{array}
\usepackage{xcolor}
\usepackage{arydshln}

\colorlet{tableheadcolor}{gray!25} % Table header colour = 25% gray
\colorlet{tablerowcolor}{gray!5} % Table row separator colour = 10% gray
\colorlet{tablerowcolor2}{gray!12} % Table row separator colour = 10% gray
\colorlet{tablerowcolor3}{gray!25} % Table row separator colour = 
\newcommand{\rowcollight}{\rowcolor{tablerowcolor2}}
\definecolor{grey}{HTML}{969696}
\definecolor{dgrey}{HTML}{01665e}
\definecolor{lgrey}{HTML}{5ab4ac}
\definecolor{improveCol}{HTML}{253494}
\definecolor{worsenCol}{HTML}{d7191c}
\definecolor{hiCol}{HTML}{d01c8b}
\definecolor{loCol}{HTML}{4dac26}
\definecolor{lgreen}{HTML}{f0f9e8}
\definecolor{dgreen}{HTML}{005a32}
\definecolor{purple}{HTML}{ae017e}

\definecolor{editCol}{HTML}{0000FF}

\newcommand{\edit}[1]{{\textcolor{black}{#1}}}

\newif{\ifhidecomments}
 \hidecommentsfalse 
\ifhidecomments
       \newcommand{\Mark}[1]{}
       \newcommand{\MarkLeft}[1]{}
       \newcommand{\MarkRight}[1]{}
       \newcommand{\Upol}[1]{}
       \newcommand{\Munmun}[1]{}
       \newcommand{\Koustuv}[1]{}
       \newcommand{\munmun}[1]{}
\else
\newcommand\Mark[1]{\textcolor{green}{\small [Mark: #1]}}
\newcommand\MarkLeft[1]{\textcolor{green}{\small $\leftarrow$[Mark: #1]}}
\newcommand\MarkRight[1]{\textcolor{green}{\small [Mark: #1$]\rightarrow$}}
\newcommand\Upol[1]{\textcolor{magenta}{[Upol: #1]}}
\newcommand\Munmun[1]{\textcolor{cyan}{[Munmun: #1]}}
\newcommand\Koustuv[1]{\textcolor{purple}{[Koustuv: #1]}}
\newcommand{\munmun}[1]{\textbf{{\textcolor{orange}{[#1 -- Munmun]}}}} 
\fi

\acmJournal{PACMHCI}
\acmYear{2023} \acmVolume{7} \acmNumber{CSCW1} \acmArticle{34} \acmMonth{4} \acmPrice{}\acmDOI{10.1145/3579467}

\setcopyright{none}

\begin{document}

% \title{Minding The Gap: How charting the Sociotechncal Gap in Explainable AI can help explainability}
\title[Charting The Sociotechnical Gap in XAI]{Charting the Sociotechnical Gap in Explainable AI: \\A Framework to Address the Gap in XAI}
% \title[The Sociotechnical Gap in Explainable AI]{The Sociotechnical Gap in Explainable AI: \\Why Bother charting It and How to Approach It}
% \author{Author Names}
% \affiliation{%
%   \institution{Anonymized}
% }

\author{Upol Ehsan}
\affiliation{%
  \institution{Georgia Institute of Technology}
  \city{Atlanta}
  \state{GA}
  \country{USA}}
\email{ehsanu@gatech.edu}

\author{Koustuv Saha}
\affiliation{%
  \institution{Microsoft Research}
  \city{Montréal}
  \state{Québec}
  \country{Canada}
}
\email{koustuvsaha@microsoft.com}

\author{Munmun De Choudhury}
\affiliation{%
  \institution{Georgia Institute of Technology}
  \city{Atlanta}
  \state{GA}
  \country{USA}}
\email{munmund@gatech.edu}

\author{Mark O. Riedl}
\affiliation{%
  \institution{Georgia Institute of Technology}
  \city{Atlanta}
  \state{GA}
  \country{USA}}
\email{riedl@cc.gatech.edu}
\renewcommand{\shortauthors}{Upol Ehsan et al.}

\begin{abstract}
Explainable AI (XAI) systems are sociotechnical in nature; thus, they are subject to the sociotechnical gap—divide between the technical affordances and the social needs. However, charting this gap is challenging. In the context of XAI, we argue that charting the gap improves our problem understanding, which can reflexively provide actionable insights to improve explainability. Utilizing two case studies in distinct domains, we empirically derive a framework that facilitates systematic charting of the sociotechnical gap by connecting AI guidelines in the context of XAI and elucidating how to use them to address the gap. We apply the framework to a third case in a new domain, showcasing its affordances. Finally, we discuss conceptual implications of the framework, share practical considerations in its operationalization, and offer guidance on transferring it to new contexts. By making conceptual and practical contributions to understanding the sociotechnical gap in XAI, the framework expands the XAI design space.

\end{abstract}


% \ccsdesc[100]{Applied computing~Law, social and behavioral sciences}


\maketitle

% \input{1introduction}

%%%%%intro start

\section{Introduction}
Explainable Artificial Intelligence (XAI) systems are \textit{sociotechnical} in nature because their technical (AI) components are embedded in social environments~\cite{ehsan2021expanding,liao2020questioning,ackerman2000intellectual}. As a research area, XAI aims to provide human-understandable justifications for a system’s behavior~\cite{ehsan2019automated,adadi2018peeking,guidotti2018survey}. Given their sociotechnical nature and the increasing trend of being deployed in high-stakes domains like healthcare~\cite{holzinger2017we,katuwal2016machine,che2016interpretable,loftus2020artificial}, finance~\cite{MacKenzie2018, murawski2019mortgage}, criminal justice~\cite{Rudin2020, Kleinberg2017, hao2019jail}, we need to account for both social and technical factors to mitigate biases and promote accountability in XAI~\cite{smith2020no}.  

All sociotechnical systems, XAI or otherwise, are subject to what \citeauthor{ackerman2000intellectual} calls the \textit{sociotechnical gap}---``the divide between what we know we must support socially and what we can support technically''~\cite{ackerman2000intellectual}.  
Ackerman posed the gap as a central challenge for sociotechnical systems, highlighting that without minding the gap, we cannot effectively design useful systems~\cite{ackerman2000intellectual}. 
One of the most intriguing and challenging aspect of the sociotechnical gap is the  difficulty to fully bridge permanently~\cite{ackerman2000intellectual, munson2013sociotechnical, dourish2011divining}. 
This is because the ``social activity is fluid and nuanced'' (which means user needs are dynamic) while the ``technical systems are rigid and brittle'', ``not socially flexible'', and ``do not allow sufficient nuance'', especially ``in their support of the social world''~\cite{ackerman2000intellectual}. 

With these challenges in mind, addressing the gap might appear futile at first: why should we bother addressing a gap that cannot be filled?
However, therein lies a crucial point made by Ackerman---the goal of introducing the gap is \textit{not} to “solve” this fundamental problem but to deeply understand it~\cite{ackerman2000intellectual,dourish2011divining}. 
Without understanding, we cannot address the gap. Without charting or mapping the gap---\textit{how} the gap looks and \textit{what} is on either side of it---we cannot understand enough to address it. The following analogy can be helpful: imagine two mountains with a canyon in the middle. If we do not know the geography of the gap formed by the canyon (e.g., where the gap is the largest or smallest), then it will be hard to find the best places to build a bridge. We need mechanisms to chart or map out the gap. This is what this paper does in the context of XAI--- provide mechanisms to chart the technical and social dimensions of XAI systems so that we can better understand and address their sociotechnical gaps. 
Charting the gap can promote greater visibility of the design space and facilitate informed design decisions. 

Charting the sociotechnical gap in XAI is challenging given AI is  ``a new and difficult design material''~\cite{ghai2021explainable, dove_ux_2017,dove_ux_2017}, especially given the non-deterministic, stochastic, and opaque-boxed nature of current systems~\cite{yang2019profiling, yang2020re}.
This implies that there is more volatility in mapping out the technical affordances of (X)AI-powered systems (vs. non-AI systems).
Thus, there is a translational challenge in charting this gap in the context of XAI systems (as opposed to non-AI systems). 

Inspired by Ackerman's goal, this paper proposes a shift in focus from a “gap filling” philosophy to a “gap understanding” one when addressing the sociotechnical gap in XAI.
We argue that \textit{engaging in the process of charting the gap can improve our problem understanding, which can reflexively provide actionable insights to improve explainability}. 
We operationalize our argument both conceptually and practically. At a conceptual level, we situate the process of charting the sociotechnical gap in XAI. By shifting the focus towards the utility of problem understanding, it can potentially highlight intellectual blinds spots in XAI, which can expand the design space. At a practical level, it can address the under-explored area of \textit{how} we might go about mapping and using it to address explainability challenges.
\textit{We address the challenges around charting the sociotechnical gap in XAI by using two real-world case studies in distinct domains to empirically derive an analytical framework to systematically chart the gap}. This framework not only integrates existing guidelines in AI and HCI but also transfers the applications in the context of XAI, highlighting the interplay between the social and technical elements.
We also explore how to connect the technical and social wings of the framework by utilizing emerging XAI concepts that take a sociotechnically-informed perspective on explainability.
We apply our developed framework to a third case study in a new domain and showcase two main outcomes: (1)~better understanding (charting) of the gap (how the gap looks) by being able to identify the most promising and challenging regions of the gap; (2)~actionable insights to address the gap (what to do with the gap) that reflexively improved informed user actionability and system explainability. In summary, our contributions are fourfold:
\vspace{-0.4em}
\begin{itemize}
\item We situate the charting of the sociotechnical gap in the context of Explainable AI, expanding the design space
\item Using two real-world case studies in different domains, we empirically derive a framework to chart the gap and apply it to a third case study in a new domain 
\item Our framework connects diverse threads of guidelines in AI and HCI, translates them to the context of XAI, elucidates how to use them to address the gap by connecting the social and technical aspects of the infrastructure
\item We discuss the conceptual implications of the framework, share practical considerations in its operationalization, and offer guidance on transferring it to new use case domains.
\end{itemize}

Our work aims to help stakeholders chart the contours of sociotechnical gaps in XAI systems. The design philosophy of our framework is generative and iterative, not normative~\cite{krish2011practical,mccormack2004generative, christians1989theory}. By ``refocusing of design in terms of process rather than solutions''~\cite{selbst2019fairness}, it can help stakeholders avoid certain techno-centric pitfalls like Solutionism (always seeking technical solutions)~\cite{morozov2013save} and Formalism (over-abstracting and seeking mathematical solutions)~\cite{green2020algorithmic}. This paper is not an exhaustive treatise of the sociotechnical gap in XAI; rather, it takes a foundational step towards addressing it by operationalizing the charting of the gap.

%%%%%intro end

%%%%%rw start
\vspace{-4pt}
\section{Background and Related Work}

\subsection{Sociotechnical Gap in HCI, CSCW, and AI systems}
\citeauthor{walker2008review} defined sociotechnical systems as the ones that include the interrelatedness of ``social'' and ``technical'' aspects. They argued that ``an inevitable consequence of mixing `socio' with `technical' is that the socio does not necessarily behave like the technical, people are not machines; paradoxically, with growing complexity, even the `technical' can start to exhibit non-linear behavior~\cite{walker2008review}.~\citeauthor{ackerman2000intellectual}, in his celebrated work~\cite{ackerman2000intellectual}, argued that there is an inherent gap between the social requirements of Computer-Supported Cooperative Work (CSCW) and its technical mechanisms, and this socio-technical gap concerns the divide between what we know we must support socially and what we can support technically. Over the years, evidence of the socio-technical gap has surfaced in various aspects and examples of Human-Computer Interaction (HCI) and CSCW technologies, such as computing affordances and data privacy~\cite{singh2007password,bazarova2015social,forte2013defining}.

AI-based computing systems are socially situated, and research recognizes the detrimental effect of a techno-centric view on AI~\cite{shneiderman2020human,vaughan20201,sabanovic2010robots}. If both social and technical factors are not considered in a balanced fashion, AI systems are hard to be integrated into individual and organizational workflows~\cite{makarius2020rising,wolf2019evaluating,saha2021life}. Further, ignoring social factors could lead to potential misuse, mistrust~\cite{yang2016investigating,yang2019unremarkable,buschek2021nine}, and ethical risks and unintended consequences~\cite{mohamed2020decolonial,suresh2019framework,sanchez2020does}. Prior work has explicitly highlighted the abstractions and traps with building AI-based computing systems in sociotechnical contexts~\cite{selbst2019fairness,andrus2020ai,morozov2013save,green2020algorithmic}. Notably,~\citeauthor{selbst2019fairness} discussed ways in which technical designers can mitigate the traps and draw abstraction boundaries to include social actors rather than purely technical ones~\cite{selbst2019fairness}.~\citeauthor{andrus2020ai} explored the many facets associated with developing AI for public interests, and how the \textit{sociotechnics} of such AI systems are closely inter-related with three contemporary areas of Fair ML, AI Safety, and Human-in-the-Loop Autonomy~\cite{andrus2020ai}.

Similarly, ~\citeauthor{alter2010design} published a conceptual paper to bridge the gap between thinking of systems as tools and thinking of systems as sociotechnical systems with human participants~\cite{alter2010design}. 
~\citeauthor{munson2013sociotechnical} drew upon Ackerman's sociotechnical gap in CSCW concepts to review contemporary sociotechnical challenges of using social media to support health~\cite{munson2013sociotechnical}.~\citeauthor{malatji2019socio} examined the socio-technical gaps within organizational information and cybersecurity practices~\cite{malatji2019socio}.

Ackerman posed the sociotechnical gap as a central challenge in CSCW~\cite{ackerman2000intellectual}. 
While there are commendable efforts to grapple with the sociotechnical challenges in XAI~\cite{liao2020questioning,ehsan2021explainable,dhanorkar_who_2021,wang2019designing,ehsan2021expanding}, addressing the gap remains understudied. 
To our knowledge, this is the first paper that explicitly focuses on charting and addressing the sociotechnical gap within the context of explainable AI systems. 
\edit{Drawing upon the rich literature in CSCW, the case studies in this paper focus on AI-supported work that is not only \textit{cooperative} but also bears \textit{collaborative} aspects (e.g., cross-functional teams). In attempting to chart the sociotechnical gap in XAI, this paper thus builds on and extends the CSCW community's long-standing tradition of designing and studying explainable computing systems~\cite{Lim2009,kulesza2013too,Eslami2015,Rader2015,schoeffer2021appropriate,haque2020understanding,lyons2021conceptualising}. 
}

\subsection{Explainable AI and its Sociotechnical Aspects}

Broadly, an AI system can be considered to be explainable (or XAI), if it aims to make its decisions \textit{easy to understand and interpret} by people~\cite{lipton2018mythos,arrieta2020explainable,gilpin2018explaining,miller2019explanation,gunning2017explainable,ras2018explanation,carvalho2019machine, ehsan2019automated}. Explainability is often viewed more broadly than model transparency~\cite{gilpin2018explaining,ras2018explanation,lipton2018mythos}. Recent XAI work showed post-hoc explanations~\cite{ehsan2019automated,lipton2018mythos} and algorithmic auditability~\cite{gilpin2018explaining} can build trust in the AI systems. Taken together, explainability is a human-factor, not just a model-inherent property~\cite{arrieta2020explainable,mohseni2018multidisciplinary,arya2019one,miller2019explanation, ehsan2020human}. Therefore, the importance of adopting user-centered approaches to XAI has been advocated in recent research~\cite{miller2019explanation,shneiderman2020human,vaughan20201,ehsan2022social}.~\citeauthor{wang2019designing} reviewed decision-making theories and identified many gaps in XAI output to support the complete cognitive processes of human reasoning~\cite{wang2019designing}. 

The HCI and CSCW communities have a long-standing interest in making computing systems explainable~\cite{lim2009and,kaur2020interpreting,amershi2019guidelines,wang2021towards,gero2020mental,luger2016like,eiband2021support}.~\citeauthor{abdul2018trends} studied HCI research on explainable systems spanning across expert systems, recommenders, context-aware technologies, and ML systems. In a series of works, Lim and colleagues have showcased the importance, opportunities, and challenges of AI intelligibility in context-aware systems~\cite{lim2009assessing,lim2009and,lim2010toolkit,lim2011investigating}.~\citeauthor{miller2019explanation} noted the need to push XAI towards human-centered instead of algorithmic-centered approaches by calling out the gaps between XAI algorithmic output and properties of explanations sought by individuals~\cite{miller2019explanation}. Further,~\citeauthor{doshi2017towards} noted the need to involve intended stakeholders in the intended usage context~\cite{doshi2017towards}.

Adopting human-centered approaches, recent works have studied explainable interfaces to help model developers diagnose and improve ML models~\cite{kaur2020interpreting,hohman2019gamut}. \citeauthor{kaur2020interpreting} examined how data scientists frequently misuse and over-trust interpretability tools~\cite{kaur2020interpreting}. Explainability features of AI systems have been explored in various domains, including health~\cite{xie2020chexplain,wang2019designing}, and recruitment/hiring~\cite{cheng2019explaining}. The value of multi-stakeholder involvement has also been noted in content moderation~\cite{lai2020chicago} and active learning~\cite{ghai2021explainable}. Complementarily, researchers have started investigating and evaluating the effectiveness of XAI techniques~\cite{dodge2019explaining,alqaraawi2020evaluating,yang2020visual,cai2019effects,buccinca2020proxy,ehsan2021explainability}. However, XAI's disconnect with the philosophical and psychological grounds of human explanations has been duly noted~\cite{mittelstadt2019explaining}, as best represented by Miller's call for leveraging insights from the Social Sciences~\cite{miller2019explanation}. Interestingly, while XAI is often claimed to be a critical step toward accountable AI, empirical studies have found little evidence that explanations improve a user's perceived accountability or control over AI systems~\cite{rader2018explanations,smith2020no}. Given explainability is a human factor and XAI systems are sociotechnical, our work expands the design space of the sociotechnical needs of XAI systems.

\subsection{Operationalizing User Needs in AI}

As AI systems get more integrated into societies, they require the best practices to be operationalized and implemented. Along these lines,~\citeauthor{amershi2019guidelines} proposed 18 design guidelines applicable for human-AI interactions~\cite{amershi2019guidelines}. There are several taxonomies of prototypical roles of XAI consumers~\cite{arrieta2020explainable,hind2019explaining,tomsett2018interpretable}, to
provide valuable insights to tackle the challenges of selecting and translating between AI algorithms. Further, researchers have studied user-centered approaches to design XAI UX~\cite{liao2020questioning,eiband2018bringing,wolf2019explainability}.~\citeauthor{liao2020questioning} studied design challenges around XAI and proposed a question-driven design process to ground the user needs, choices of XAI techniques, design, and evaluation of XAI UX~\cite{liao2020questioning}. Recently,~\citeauthor{ehsan2021expanding} expanded the notion of explainability in AI systems by introducing and exploring the concept of {\em Social Transparency}, ``a sociotechnically informed perspective that incorporates the socio-organizational context into explaining AI-mediated decision-making.''~\cite{ehsan2021expanding}. The authors examined how socio-organizational information carried by the 4W---\textit{who} did \textit{what, when}, and \textit{why}---design features can augment AI explainability~\cite{ehsan2021expanding}. Towards addressing the societal issues of algorithmic governance, \citeauthor{lee2019webuildai} presented a collective participatory framework that enables people to build
an algorithmic policy for their communities~\cite{lee2019webuildai}.

Prior work has also advocated or proposed means to operationalize AI systems to be more human-centered and socio-organizationally situated~\cite{choudhury2020introduction}.~\citeauthor{ehsan2020human} adopted Critical Technical Practice~\cite{agre1997computation} and Reflective Design~\cite{sengers2005reflective} to propose the foundations of a reflective Human-Centered XAI (HCXAI). 
\citeauthor{madaio_co-designing_2020} took a participatory approach to co-designing checklists to understand organizational challenges are AI ethics~\cite{madaio_co-designing_2020}.
and~\citeauthor{zhu2018value} proposed Value Sensitive Algorithm Design (VSD)~\cite{zhu2018value} by engaging stakeholders in the early stages of algorithm creation to avoid biases in design choices or compromising stakeholder values~\cite{zhu2018value}. Taking an ``AI lifecycle'' perspective, \citeauthor{dhanorkar_who_2021} underscore how the explainability needs of different types of explanation audiences evolve over time~\cite{dhanorkar_who_2021}. Prior work has also leveraged design fiction and speculative scenarios to understand user values and cultural perspectives for AI system design~\cite{cheon2016integrating,cheon2018futuristic,muller2017exploring}.~\citeauthor{sabanovic2010robots} proposed a framework of Mutual-Shaping and Co-production~\cite{sabanovic2010robots}.~\citeauthor{jones2013design} proposed a design process for intelligent sociotechnical systems with equal attention to analysis of social concepts in the deployment context and representing such concepts in computational forms. On the side of technical affordances,~\citeauthor{sanneman2020situation} designed a three-level framework for the development and evaluation of an explainable AI system's behavior~\cite{sanneman2020situation}.

In other works,~\citeauthor{gebru2018datasheets} proposed \textit{datasheets for datasets}, aimed at facilitating better communication between dataset creators and dataset consumers, and encourage the prioritization of transparency and accountability~\cite{gebru2018datasheets}. \citeauthor{sokol2020explainability} proposed explainability fact sheets to enable researchers and practitioners to grasp the capabilities and limitations of a particular explainable AI method~\cite{sokol2020explainability}. On the side of models,~\citeauthor{mitchell2019model} proposed model cards to encourage transparent model reporting around creation, curation, and evaluation~\cite{mitchell2019model}. On trust and AI ethics,~\citeauthor{hind2018increasing,bellamy2018ai} have explored operationalizing it in organization settings~\cite{hind2018increasing, bellamy2018ai,knowles2021sanction}. 

Our work builds upon and contributes to the above body of work to propose a framework that can help identify, operationalize, and address the gaps in various social and technical components in explainable AI systems and AI explainability. For this, we are not only informed by the above perspectives but also integrate them to chart the sociotechnical gap in explainable AI systems and bridge it at an infrastructural level.

%%%%%rw end

%%%%%cs start
\section{Case Studies on Real-world Applications of XAI} \label{section:casestudies}
We share two case studies from different domains: sales (Section~\ref{sec:kuro}) and mental health (Section~\ref{sec:clinical}). These case studies represent real-world projects spanning two years of engagement by one or more of the authors. For each case study, a background section provides necessary context, the problem, descriptions of the interplay between the social and technical components (which lead to scoping the gap), and steps we took to address the mapped out sociotechnical gap. While outlining the interplay, we focus on relevant aspects which ultimately help construct a framework (Section~\ref{sec:details_framework}) that helps us map out the sociotechnical gaps in new use XAI use cases. Providing an exhaustive list of technical attributes and social needs is neither productive, nor necessary, nor potentially possible (the list might be uncountably large). 

Instead of offering a top-down narration of providing the framework first, it may be helpful to the reader to take a bottom-up "build-up" approach to build up the contextually relevant parts first and allows us to link back to the concepts later. 
This approach is also representative of the real-life events of how we arrived at the framework. 
We ground our data and motivation for the building blocks of the framework through the first case study and use those lenses to describe the second.
The structuring can be potentially helpful to the reader for transferring the process to other cases.
We structure the interplay in a way that serves as building blocks for charting the sociotechnical gap and the eventual development of the framework. 
\edit{We not only share the two case studies to empirically derive the framework but also apply the framework through another real-world case study in a different domain (cybersecurity) (\autoref{sec:application_framework}). We offer an end-to-end, conception-to-implementation, narrative of how charting the gap can help understand and address it. We also discuss how the framework can be transferred to other domains in~\autoref{sec:implications}
}

\subsection{Case Study 1: Kuro, the AI-powered sales software}
\label{sec:kuro}

\subsubsection{Background} SalesCorp (pseudonym), a Fortune 100 multi-national technology company, invested US \$5 million to build an AI-powered pricing recommendation system to help its technology salesforce. Each sale typically ranges from US \$500K to \$2M;
therefore, the stakes are high. If a client is priced out of their budget, SalesCorp might lose the immediate sale and risk losing long-term revenue from the client. SalesCorp follows a business-to-business (B2B) model and sells hardware, software, and cloud infrastructures. It follows a dynamic pricing model, dependent on multiple factors like a client's historical budget, company size, length of the relationship, etc.

The AI system, Kuro (pseudonym), was introduced as a ``virtual experienced peer who is always there'' and claims to utilize over 50,000 data points to drive recommended prices. The system was developed in a 70-30 partnership, 70\% developed in-house using proprietary data and 30\% developed with third-party vendors that offer AI infrastructural support (e.g., proprietary models licensed for exclusive use). A promised value proposition of AI-powered Kuro was its robustness, an alleged improvement from its ``brittle'' predecessor Sumo.

\subsubsection{The Problem} On the surface, Kuro’s metrics were impressive: 90\% accuracy, good model transparency (showing top 3 features), and confidence range (0-100\%) shown next to each recommendation. 
However, there was a breakdown --- only 10\% of the sellers used Kuro; of those who used it, only 2\% found it useful. Given the socio-organizational context of the deployment, the problem here was not a purely technical one; it had strong social components (as we will see below).
Considering Kuro’s problems, SalesCorp recruited an author of this paper to work on this issue.

\edit{Chronicling a year-long collaboration, below we motivate and ground our methods and the data that ultimately form the buildings blocks the \textit{social} and \textit{technical} wings of the framework. 
Note that we structure the presentation of the following events to help the reader find correspondence between the case studies below and the framework derived from it (in the next section).
Even though we structure the two prongs sequentially in our write-up, they took place in a largely parallel fashion,
iteratively building on top of each other to provide a clearer picture of the sociotechnical gap. To protect client confidentiality, proprietary details are redacted from quotes.}

\subsubsection{\edit{Grounding the Data, Process, \& Motivations for the Building Blocks for the Framework}}
\label{sec:framework_derivation}
\edit{Recognizing the socio the sociotechnical nature of the infrastructure, we focused on two relevant areas of the sociotechnical gap: understanding \textit{the technical affordances} and \textit{the socio-organizational} needs. To understand these two aspects, we took the following approach.}

\edit{Taking a participatory approach~\cite{schuler1993participatory}, we conducted 2 workshops with 28 participants each. 
Participants had diverse roles (5 XAI Engineers, 6 Product Managers, 5 Responsible AI Managers, 3 Data Scientists, 4 VPs of Technology, and 5 UX Researcers). 
We took a participatory approach to leverage the domain expertise of our stakeholders and have their voices actively represented in our interventions.   
During orientation, we introduced the concept of the {\em sociotechnical gap} to our stakeholders so that we can focus on the divide between what the current system supports technically and what we must support socially. 
Almost half of them were already aware of the concept of the sociotechnical gap. 
Moreover, using examples of related work~\cite{dhanorkar_who_2021,ehsan2021expanding,liao_question-driven_2021,liao2021human}, we oriented them to a human-centered XAI lens, one that does not restrict XAI to merely model transparency and incorporates the sociotechnical factors in its conceptualization~\cite{ehsan2020human}. This helped the participants adopt a sociotechnical stance (vs. a techno-centric one) in problem-solving.}

\edit{Below we share \textbf{key takeaways} from each workshop to ground our data and motivate the structure of the framework.}

\vspace{0.5em}
\edit{\textit{\textbf{Workshop 1: constructing building blocks.}} The goal in this 3-hour workshop was to \textit{brainstorm and reach consensus on the relevant building blocks to consider when thinking about technical and social sides of the problem} (later forming the foundations of the resulting framework in Sec~\ref{sec:details_framework}).}

\edit{The first half of the workshop consisted of a generative (brainstorming) exercise where four cross-functional teams used sticky notes to list out the relevant topics (building blocks). In this stage, we explicitly asked our participants to focus on the generation of ideas and not an evaluation of them. The teams generated 7 blocks under the technical wing and 12 under the social side.} 

\edit{In the second  half, we concentrated on narrowing down the building blocks in a way that balances practical feasibility with conceptual rigor. 
All four teams reported that a total of 19 blocks were too many to be practically useful.
Next, we paired up the teams --- four teams became two; once each of the two teams reached a consensus, two teams became one large team.
Given the cross-functional nature of each team, the discussion challenged assumptions in a positive way relayed by this data scientist, ``It’s pretty refreshing to have your views challenged. I never really considered why I need to know the social aspects. This process helps me see exactly why I was tunnel-visioned.'' 
An iterative affinity diagramming process generated three thematic blocks for the technical side and three for the social side (we did not put any requirement for balancing the number of blocks).}

\edit{The final task was to figure out the names for these blocks. 
We asked them to share the top function of each block. 
On the \textbf{technical} side, they listed: \textbf{data} genealogy (where the data comes from, how was it collected, etc.), \textbf{model}/algorithmic affordances (what can or cannot the model do), and how the AI system generates \textbf{explanations}. 
On the \textbf{social} side, they outlined: \textbf{trust} (how much or how little users rely on the AI), \textbf{actionability} of the explanations (what users need to act on the AI-generated explanations), and organizational \textbf{values} (how organizational incentives and values interplay with individual expectations). 
Thus, the group settled on the final blocks—on the technical side, there were data, models, and explanations; on the social side, we had trust, actionability, and values. We use these building blocks as analytic lenses to structure the case studies and also as foundations for the framework (\autoref{sec:details_framework}).}

\vspace{0.5em}
\edit{\textit{\textbf{Workshop 2: what is inside each building block.}} 
In this 4-hour workshop, the goal was to \textit{operationalize each block in a systematic manner}. What questions or guidelines may we use to practically explore each block that will increase our understanding of the sociotechnical gap? 
The first step was to collaboratively explore sets of questions or guidelines we can use to flesh out the idea behind each block. To achieve this goal, initially, we proposed custom-made questionnaires (that we developed) for each block. However, we received substantial push-back against introducing new materials without adequately leveraging existing workflows and processes. For instance, the company had existing processes where engineers and data science teams prepared paradata documentation like Datasheets for Datasets~\cite{gebru2018datasheets}. Citing adoption concerns for new questionnaires, one product manager critically asked "who will say yes to doing new work without exploring how far we can go with what we have?" Building on this feedback, we pivoted into leveraging existing processes and documentation. Similar to the first workshop, we split participants into four cross-functional teams with the common mandate of finding existing processes and guidelines relevant to different building blocks. Next, we paired up the teams to facilitate the iterative filtering till a consensus was reached (four teams became two and two became one). The teams were able to highlight important. For instance, teams agreed that we could use Model Cards~\cite{mitchell2019model} for the model block and Datasheets for Datasets for the Data block.
}

\edit{The second step was to decide how many questions or guidelines should go under each block and how we may cluster them. 
To balance practical viability and burden, we received feedback to \textit{not} generate an exhaustive list of questions. As a data scientist put it: “give us \textit{starter questions} [from different guidelines] that kickstarts the conversation… [and] allow us to filter questions from the remaining set” (emphasis added). 
For each block, using thematic analysis~\cite{braun2006using}, we worked with participants to incorporate relevant “starter” questions from different guidelines. 
Next, we used affinity diagramming~\cite{beyer1996contextual,bella2012universal} with stakeholders to cluster relevant questions for each topic under each wing (e.g., which ones from Datasheets for Datasets are most relevant).
Last, we incorporated the feedback of 6 XAI and HCI experts to refine the final list of starter questions.}

\edit{This grounding forms the foundation of the resulting framework (\autoref{sec:details_framework}). Equipped with this understanding, we now share details on how we explored the technical and socio-organizational affordances to solve the problem with Kuro.}

\subsubsection{Understanding the technical affordances} 
There are three building blocks to unpack the technical wing: \textit{data}, \textit{model}, and \textit{explainability}.

For \textbf{data}, using a participatory approach, we appropriated the questionnaires from Datasheets for Datasets~\cite{gebru2018datasheets} and customized it for the company, capturing insights on the data genealogies. 

For the \textbf{model}, we hit ``off limits'' spots because the vendors supplied some of the decision mechanisms. When possible, we conducted interviews with account managers from the vendors and adapted Model Cards~\cite{mitchell2019model} to our use case. While we could not exhaust each entry from the original template, the empty slots were just as useful as the ones with data. It was an explicit reminder of what we do not have, which ultimately helps us scope the technical affordances.

For \textbf{explainability}, we leaned on two main artifacts – first, we selected relevant questions from~\citeauthor{liao2020questioning}’s XAI Question Bank~\cite{liao2020questioning}. Combined with the Datasheets and Model Cards, the questions provided a good starting point to explore different facets such as explanation type (how, why, what-if, why not, etc.). Second, we adapted IBM's Explainability Fact Sheets~\cite{bellamy2018ai} to get a sense of the explanation generation mechanism (modality, global vs. local, etc.) at functional and validation levels. We had to use a subset because the original Fact Sheets had a list 36-item worklist. As one developer put it, “it was too much trouble for too little value.”

\subsubsection{Understanding the socio-organizational needs} The social side happened in tandem with the technical side. 
We want to get clear feedback on why the engagement was at 10\%. To unpack this problem, we used a combination of interviews, Envisioning Card exercises~\cite{friedman2012envisioning}, and workshops to understand three main things: trust in AI (what is harming? What is helping? What else is needed?), actionability of the explanations (what could help users take informed actions on the AI’s decision), and values (tensions and alignments with organizational norms). 
When investigating \textbf{trust}, we learned that while the model transparency was helpful, it was not enough to calibrate the trust in AI. Less experienced sellers tended to see the multi-million-dollar Kuro as “godlike”. They tended to over-trust the AI and felt they lacked the voice/agency to disagree with the AI. Instead of putting them in that position, they chose to completely ignore/not engage with the AI. More experienced sellers simply ignored the AI’s recommendation because they felt they “knew better”. Understanding the nuances of trust helped us scope the social needs better. 

Thinking of \textbf{actionability}, Kuro’s explanation information was deemed ``inert''. One said, ``I’ve no idea of what to do with the numbers [Kuro] shows me. How is it confident? Confidence with respect to what?'' Moreover, there was a mistake in the assumed interaction paradigm—it was  1 Human -1 AI, whereas the activity of selling is a group one (many-to-many). Many sellers worked around this siloed constraint by using analog means. They would walk over to other members of the team (each client was typically served by a team of 5 sellers of differing hierarchies) to consult.

When considering \textbf{values}, we found multiple avenues of tensions in organizational norms. Overriding Kuro’s decision was extremely cumbersome—the audits and escalations involved created a disproportionate burden on the sellers, which is why they consistently ignored it. Moreover, entry-level sellers had no idea of what ``acceptable'' changes to Kuro’s prices are. The sellers perceived the technical staff as “invaders”, ones who want to automate the sellers out of their jobs. This created a tense relationship, hindering collaboration between the stakeholders.

\subsubsection{Addressing the gap} Compared to when we started, these processes painted a clearer picture of the two sides of the sociotechnical gap and highlighted opportunities to use design as a bridge. While we acknowledged that the gap would be ever-present, the goal was never to “fill” it, but to understand the boundaries and act on it through what \citeauthor{ackerman2000intellectual} calls first-order approximations---``tractable solutions that partially solve specific problems with known trade-offs''~\cite{ackerman2000intellectual}. 

A core takeaway that connects trust, actionability, and value tensions was sellers' desire to know how others acted on the AI’s recommendation. They wanted “peripheral vision” of past user trajectories. For example, a junior seller would want to see what a senior person did with the AI system and why. This additional context, the seller felt, would help them judge the AI better, making the explanations actionable. 

We tested if adding socio-organizational context would indeed help. In an ad-hoc manner, we offered a piece-meal approach. We asked the sellers to pin a Google Sheet (online collaborative spreadsheets) in their slack channel and put the following information: for any recommendation from Kuro that was borderline, they would notate the incident ID, what they did (accept/reject), and write a Tweet-length justification (sellers appreciated the ``Tweetification'' of the ``why''). Within one week, the Google Sheet had over 400 entries. Sellers even appropriated the original format by adding “+1” (like) voting to useful comments. We saw the engagement rise to 42\% (from 10\%). This was particularly surprising because the sellers went out of their workflow to engage with the Google Sheets. Interviews showed that they found enough value in this ``peripheral vision'' to make it worth overcoming the context-switching burden. 

With a proof of concept in place, we could propose engineering changes. 
Through participatory design activities, we designed an intervention where next to Kuro’s AI recommendation and technical transparency, the seller would be able to see \textit{who else did what, when}, and \textit{why}. 
While developing the intervention, we followed the framework for Social Transparency~\cite{ehsan2021expanding} to foster calibration of trust in AI and consistency in collective actions. 

After implementation, we carefully monitored engagement, which had improved to 87\%, a remarkable jump from 10\%. Moreover, sellers shared that they understood Kuro better because the AI's decision was now in context of human actions. 
The additional context cultivated a holistic form of explainability, which was not possible purely through technical transparency. Through follow-up interviews, surveys, and workshops, our conception of the boundaries of the gap got refined, which brought previously unconscious aspects into conscious awareness, thereby making it actionable. Our actions reflexively made Kuro’s explanations actionable, improving their explainability.

\subsection{Case Study 2: An AI-Powered Clinical Mental Health Tool}
\label{sec:clinical}

\subsubsection{Background}
The global burden of mental illnesses accounts for 32\% of years lived with
disability, making these conditions a major contributor to the global burden of disease~\cite{vigo2016estimating}. 
While timely treatments are useful, mental heath treatment heavily relies on \textit{what} the patient tells their clinician during in-person consultations. However, clinicians have often reported experiencing tension between a patient's self-report and external reality~\cite{american2015american}. 
\citet{fisher_beyond_2017} 
provided case reports where clinicians had begun to incorporate patients' electronic communications, such as social media, as a new form of collateral information. 
Prior work in social computing and HCI has shown that, from such data, AI and machine learning can help to infer whether an individual is vulnerable to mental health conditions~\cite{saha2019social,saha2022social,chancellor2020methods,guntuku2017detecting}, such as major depression~\cite{de_choudhury_predicting_2013} and  postpartum depression~\cite{de_choudhury_characterizing_2014}.
An emergent body of work in digital mental health has also argued how the computational power harnessed by AI systems could be leveraged to reveal the complex psychopathology of psychiatric disorders and thus better inform therapeutic applications and collaboration across different clinicians~\cite{yoo2020designing,lee2021artificial}. 

\subsubsection{The Problem}

Although many patients already bring social media data to their appointments with clinicians and patients~\cite{balick2014technology}, currently, there is no technology to support these interactions, especially given that different members of the clinician team would have different needs to fulfill the same types of data. 
To support realizing the potential of social media-derived AI insights at the point of care, in this case study, clinical researchers and practitioners from a large, urban, nationally-recognized health system in the Northeast of the U.S. collaborated with one of the authors of the current paper~\cite{yoo2020designing,yoo2021clinician}. Their \textit{central goal} was to investigate if AI, when applied to these data voluntarily contributed by patients, could provide different faceted views (e.g., through a dashboard) that would support the varying needs of different treatment team members. 

A prototype of the tool, developed via a co-design approach through a year-long collaboration and by harnessing voluntarily shared social media (Facebook) data of consented mental health patients, was tested with various types of mental health clinicians at this large health system through 1-1 interviews (\textit{N}=13) and focus groups (\textit{N}=13).
This case study revealed promise for the prototype but also highlighted issues such as liability concerns around using a tool that was a ``black-box'' to the clinicians. They imagined scenarios where despite access to patient data that indicates an exacerbation of symptoms, clinicians might not be in a position to take action. Like with the previous case study, we outline the \textit{social} and \textit{technical} parts of the \textit{infrastructure}.

\subsubsection{Understanding the technical affordances}

The interviews and focus groups revealed a variety of technical features that the clinicians deemed important from the perspective of supporting collaborative treatment to their patients. Following the style adopted for the previous case study, we discuss these technical affordances along the same four dimensions of \textit{data}, \textit{model}, and \textit{explainability}. 

Focusing on the \textbf{data}, the study highlighted important issues around mental health symptoms and clinical risk factors. The prototype was deemed promising to the clinician participants; however, they desired that the tool needs to facilitate collectively identifying and visualizing the range of different markers of the patient's mental health state that might be differentially valuable to different clinicians, as a patient navigates their interactions with psychiatrists, therapists, and social workers. 

For the \textbf{model}, participants noted credibility issues regarding the AI-derived mental health insights the prototype provided. There were two dominant credibility issues that could diminish trust: if social media data correlates to actual life (construct validity) and if an AI applied on top of this data can distinguish different contexts and intents behind specific posts (clinical utility). 

Finally, turning to \textbf{explainability}, the participants said that the tool can be useful to compare and negotiate what the social media data shows and what the patient verbally reports, especially if the two are mutually conflicting. It can also help one clinician understand how and why another member of the team appropriated certain social media-derived AI insights but may have chosen to ignore others. These technical needs necessitate incorporating explainability features in the tool.

\subsubsection{Understanding the socio-organizational needs}

In tandem with the technical needs, the interviews and focus groups also discovered a range of social and organizational needs for the AI-based tool to be successful at the point of care. We discuss these below.

The clinician participants raised a number of concerns surrounding \textbf{trust} in use of the tool. 
In the absence of an implicit level of trust in the functioning of the algorithms, clinician participants felt that they would consider the social media based AI insights with caution. Importantly, participants were concerned that trust in the tool's abilities and clinical usefulness might be undermined by the observer effect~\cite{de2013observer,saha2023observer}, wherein patients may stop posting or begin to self-censor themselves on social media, knowing clinicians' awareness of and access to this information. 

Speaking of \textbf{actionability}, 
different clinicians may also act upon the presented information in the tool in different ways: a therapist could use this data as a complement to reports from family members since the supposed ``objectivity'' of the AI insights could resolve and reconcile different collateral sources to construct a more accurate picture of the patient’s status. In addition, some participants speculated if the tool could play some preventive role 
where it allowed comparing a particular pattern of symptoms (say, suicidal thoughts) prior to hospitalization with current patterns, this could be useful for engaging preventive crisis intervention services.

Lastly, a number of conflicts in \textbf{values}, whether between the patient and the clinician, or different members of the clinician team were found. 
With regard to clinician-clinician value tensions, the case study unraveled underlying issues of burden. Participants were worried about the potential burden on their work hours and expectations in the use of the tool, perhaps more for some types of clinicians than others. Others felt that it may not always be feasible to review patients’ social media information prior to consultations because patient loads can be exceedingly high. Overburdened clinicians (e.g., social workers) often adopt a work style and value systems distinct from clinicians who provide medication assistance to a handful of patients.

\subsubsection{Addressing the gap}
We now outline some possible interventions to mind the sociotechnical gap in the context of this case study. Like the first case study, the goal of these interventions is to understand the boundaries of the technical and the social needs~\cite{ackerman2000intellectual}, rather than seeking to ``fill'' it.

First, many clinical teams schedule synchronous meetings periodically to engage in peer-review of treatment protocols and discuss discrepancies in the standard of care. If we integrate AI-driven insight from patients' social media data, we would need a way to summarize, synthesize, and visualize them for later use. To facilitate collaboration, we can envision future iterations that integrate these AI-driven insights with meeting notes into the Electronic Health Record (EHR) for later review. 
Together, these adaptations to the design of the tool can facilitate bringing not only transparency to these discussions but also improving the actionability of the AI. 

Second, we learned that the processes underpinning \textit{how} different clinicians acted on the AI insights in the prototype were missing. For instance, if a trend shows exacerbated symptoms and a therapist sees this information was not used by the psychiatrist, this information gap 
can hamper the therapist's treatment plan, create confusion, and diminish trust in the tool.
The clinician participants therefore strongly felt that \textit{transparency to support these social interactions} will be needed in the future.
For instance, we can design filtered views that showcase the decision-making pathways relevant to the specific patient based on their specific symptoms or risk factors.

%%%%%cs end

%%%%%fr start
\section{A Framework to Chart the Sociotechnical Gap in XAI}
\label{sec:details_framework}

Here we share our design lens in XAI, the framework's  setup, derivation, and its building blocks.

\vspace{-1em}

\edit{
\subsection{Our Design Lens in XAI: Human-Centered Explainable AI}
Given our goal is to chart the sociotechnical gap in XAI, we adopt a design lens in XAI that is sociotechnically-informed—that of Human-centered XAI (HCXAI)~\cite{liao2021human,ehsan2020human,dhanorkar_who_2021}.
HCXAI expands the concept of explainability beyond the bounds of the algorithm~\cite{ehsan2021expanding} and positions it as a relational and audience-dependent construct instead of a model-inherent one~\cite{arrieta2020explainable,mohseni2018multidisciplinary,arya2019one,miller2019explanation}. 
Given AI systems (Human-AI assemblages~\cite{ehsan2021expanding,dhanorkar_who_2021}) exist in sociotechnical settings~\cite{sun2022investigating, liao_question-driven_2021}, it takes more than just algorithmic transparency to make them explainable~\cite{ehsan2022human,miller2019explanation}. 
Thus, explaining what is happening “inside the black box” often requires us to also understand things “outside the black box” ~\cite{dhanorkar_who_2021, liao2020questioning}, requiring us to consider the entire AI lifecycle (vs. just the algorithm). 
For instance, why a facial recognition system disproportionately misclassified women of color~\cite{buolamwini2018gender} can be explained by looking at demographic compositions in the training data.
Emerging work in HCXAI ~\cite{schoeffer2021appropriate, haque2020understanding, pushkarna2022data} showcases how a broader XAI perspective can potentially address criticisms of popular algorithm-centered XAI techniques, which can be ineffective~\cite{alqaraawi2020evaluating,poursabzi2018manipulating,zhang2020effect} and potentially risky~\cite{kaur2020interpreting,stumpf2016explanations}.
}

 \vspace{-1em}
\subsection{Overall Setup to Derive the Framework}
\begin{figure}[t]
    \centering
     \includegraphics[width=\columnwidth]{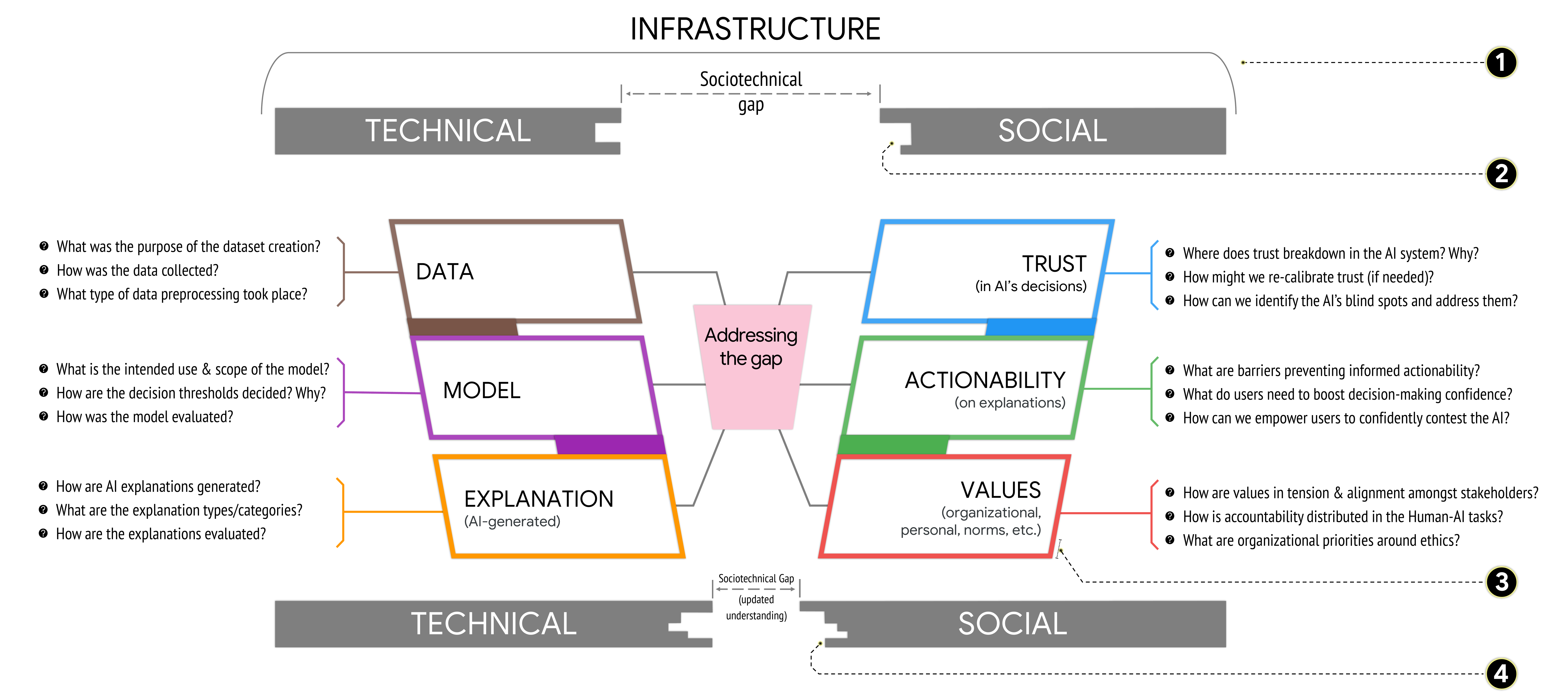}
     \vspace{-0.5em}
    % \includegraphics[width=13.1cm]{images/violinsNonCS_bw.png}
    % \begin{fignote}
    % Note: 
  \caption{
    Depicting the framework and its effect on the sociotechnical gap (in an XAI context) \textit{before} and \textit{after} the process. \textbf{(1)} \textit{At the top},  we begin with the infrastructure consisting of both social and technical wings (grey boxes) with the sociotechnical gap in the middle. \textbf{(2)} Before charting the gap, we have a “rudimentary” idea of \textit{how} its boundaries look like, depicted using the jagged edges. \textit{In the middle}, we engage in the process of charting the gap with a “look inside” each wing by operationalizing the building blocks. The sequence of the blocks do not signify or impose any sequential limitations. As we will see during framework application in Sec.~\ref{sec:application_framework}, the sequence of traversing the blocks is dependent on the problem and insights we gain from each block. \textbf{(3)} The slant in each wing is meant to convey that by engaging in the charting process, we gain actionable insights that address the gap (visually, bringing the two wings closer together). \textbf{(4)} Finally, \textit{at the bottom}, having gone through the charting process, we emerge with a ``higher resolution'' understanding of the gap visually represented by the more fine-grained jagged edges of technical and social wings (different than the top grey boxes). 
    The mapped out view reveals points where the gap is closer (highlighting promising areas to start addressing) and farther (challenging areas) from each other. \textit{Overall}, the top represents a rudimentary idea of the gap before the mapping process, the middle signifies a “look inside” each half during the process, and the bottom depicts a refined and well-charted gap. Example “starter questions” are provided alongside each block that can jump-start its operationalization.
    } 
    % \textit{more caption items to follow}}
    % \end{fignote}
            \label{fig:framwork}
    % \Description[TBD]{TDB}
    \Description[figure]{A figure to demonstrate the framework charting the sociotechnical gap in XAI systems. On the left side (technical wing), there are the blocks of Data, Model, and Explanation, and on the right side (social wing), there are Trust, Accountability, and Values. A trapezoidal block of Social Transparency bridges the social and technical wings. The top of the figure represents a rudimentary idea of the gap before the charting process, the middle signifies a “look inside” each half during the process, and the bottom depicts a refined and well-charted gap.}
    \vspace{-1em}
\end{figure}

We shared the data grounding along with the process and motivation behind the building blocks in~\autoref{section:casestudies}.
The framework begins by explicitly acknowledging the sociotechnical gap through two wings: the \textbf{technical} and the \textbf{social}. 
Figure~\ref{fig:framwork} depicts the idea---each wing has three interconnected building blocks. On the \textit{technical wing} are Data, Model, and AI-generated Explanations. On the \textit{social side}, the building blocks are Trust in AI, Actionability of Explanations, and Values. To operationalize the goals for each block, we share plausible stakeholders (Table~\ref{table:framework}) at each step and offer “starter packets” with methodological recommendations and insights from existing guidelines. The starter questions provided next to each block in Fig.~\ref{fig:framwork} should not be construed as being comprehensive because there is no one-size-fits-all formula in real-world AI deployments---the context, domain, and target audience govern how the process manifests. 
Example questions are meant to inspire and jump-start the generative process of charting the gap. 
The questions for each block were collaboratively scoped from existing guidelines (highlighted above) through our case studies.
As Ackerman pointed out, the dynamic nature of user needs (social side) is primarily why the sociotechnical gap is hard to fully bridge~\cite{ackerman2000intellectual,dourish2011divining,munson2013sociotechnical}. Thus, problems often arise from the social side such as due to value tensions with AI's recommendation, users have difficulty acting on it.

Below we describe the operationalization of each building block, starting with the technical wing and followed by the social wing. 
As we highlighted in our case studies (in Section~\ref{section:casestudies}), even though we structure the two wings sequentially in our write-up, they are interconnected and iteratively build on top of each other to provide a clearer picture of the sociotechnical gap.
The order in which the blocks appear (in the diagram or our writing) has no bearing on the order in which they can be operationalized (we can go back and forth, switch wings, etc.). The sequence of block traversal depends on the problem and findings at each stage; for instance, the framework application case study (in Section~\ref{sec:application_framework}) will showcase how the process can start on the Actionability block (the social wing) and then move to the Data block (the technical wing). 
After describing the building blocks, we provide guidance on how to move around them using thematic labels (tags) that can keep track of the iterative process. 
Next, in terms of addressing the gap, we share strategies in the form of\textit{palliatives} and \textit{first-order approximations}~\cite{ackerman2000intellectual} to address the gap. 
Finally, we share what end-products one might expect from engaging in this process. 
Beyond the mapping and charting affordances of the framework, a core contribution lies in integrating existing AI and XAI guidelines in the context of the sociotechnical gap. 

\begin{table}[t]
\sffamily
\footnotesize
 \caption{Summary of the framework to chart the sociotechnical gap in XAI.}
\begin{tabular}{p{0.13\columnwidth}p{0.28\columnwidth}p{0.24\columnwidth}p{0.24\columnwidth}} 
\textbf{Building Block} &  \textbf{Goal} & \textbf{Operationalization Examples} & \textbf{Example Stakeholders}\\
\toprule
\rowcollight\multicolumn{4}{l}{\textbf{Technical Wing}}\\
Data & To understand data genealogy, affordances, and scope what the data can tell us. & Datasheets for Datasets~\cite{gebru2018datasheets}, AI FactSheets 360~\cite{arnold2019factsheets}, Dataset Nutrition Label~\cite{holland2020dataset} & Data Scientists, Data Analysts, Database Engineers\\
\hdashline
Model & To scope the model’s affordances and limitations, which are important for trustworthiness. & Model Cards~\cite{mitchell2019model} & Data Scientists, AI Operations Engineers, Product Managers\\
\hdashline
Explanations (AI-generated)& To scope the affordances and limitations of AI-generated explanations especially around generation techniques, modality, and category. & XAI Question Bank~\cite{liao2020questioning} & Data Scientists, XAI Engineers, End-users\\
\rowcollight\multicolumn{4}{l}{\textbf{Social Wing}}\\
Trust (in AI) & To understand the current level of trust and to scope calibrating trust. & Surveys~\cite{hoff2015trust,siau2018building,ashoori2019ai}, Interviews guided by the starter question, and Value Sensitive Design (VSD) exercises~\cite{friedman2002value} & End-users including sales personnel and account managers\\
\hdashline
Actionability (of Explanations) & To understand what users need to act on the AI’s explanation in an informed manner. & Social Transparency Framework~\cite{ehsan2021expanding} & End-users of the system (e.g., analysts or mental health specialists)\\
\hdashline
Values & To understand organizational values and their interplay with individual expectations. & VSD exercises, participatory design workshops & Multi-stakeholder involvement, including leadership, HR, developers, and end-users.\\
\bottomrule
\end{tabular}
\label{table:framework}
\Description[table]{A table showing a summary of the framework to chart the sociotechnical gap in XAI. The table has four columns and eight rows. The title of the columns are Building Block, Goal, Operationalization, and Example Stakeholders. The first row is a sub-header of Technical Wing, and the next three rows are about its three building blocks, Data, Model, and Explanations (AI-generated). The fifth row is a sub-header of Social Wing, and the next three rows are its three building blocks, Trust (in AI), Actionability (of Explanations), and Values.}
\vspace{-6pt}
\end{table}

\subsection{The Technical wing}
This wing of the framework aims to scope out what the system can afford and also cannot afford. We partition this space into \textit{three} interconnected blocks: \textbf{data, model,} and (AI-generated) \textbf{explanations}. The motivation for this partition is two folds: 
first, the partition is theoretically situated in the spectrum of the technical infrastructure for XAI systems utilized by existing guidelines (e.g., Model Cards~\cite{mitchell2019model}, AI Fairness 360~\cite{bellamy2018ai}, XAI Question Bank~\cite{liao2020questioning}, etc.). Second, it is empirically grounded in our case studies where we found that such a partition struck a balance between analytical demarcation and practical operationalization that resonated with stakeholders. This alignment promoted transferability from related guidelines, which facilitated the operationalization of each building block (detailed below).

\vspace{0.5em}
\textit{\textbf{Data:}}
The goal of this building block is to understand data genealogy, affordances, and scope what the data can or cannot tell us (to the best extent available or possible). Data infrastructure is the food for the model, which entails it has fundamental implications on the entire lifecycle. Thus, we need a reliable and situated nutrition label for the data. To \textit{operationalize} the goal, we can lean co-opt existing guidelines such as Datasheets for Datasets~\cite{gebru2018datasheets} and AI FactSheets 360~\cite{arnold2019factsheets}, and Dataset Nutrition Label~\cite{holland2020dataset}. Utilizing existing guidelines and adapting when needed, we are broadly interested in the origin story of the dataset. Fig.~\ref{fig:framwork} shows some of the starter questions we can use to understand the purpose of the creation, duration of the data collection, types of preprocessing, how recent the dataset is, sampling information, etc. The starter questions often inspire follow-up questions that are more context-specific. It is important to balance breadth and depth when getting the data genealogy. The \textit{stakeholders} of this process are variable across organizations but typically tend to be data scientists, data analysts, database engineers, etc.

\vspace{0.5em}
\textit{\textbf{Model:}}
This block aims to scope the model's affordances and limitations, which are important for trustworthiness. Models are entities that transform the data into something of value (predictions, classifications, etc.). To \textit{operationalize} the goal, we can co-opt guidelines like Model Cards~\cite{mitchell2019model}. Fig.~\ref{fig:framwork} provides some starter questions around the model's intended purpose, architecture, performance evaluation, and datasets for evaluation. The starter questions are designed to seed deep generative discussions around specific use cases. The \textit{stakeholders} are variable but typically tend to be data scientists, AI operations engineers (ones who validate and test models), product managers, etc. 

\vspace{0.5em}
\textit{\textbf{Explanation:}}
Here, we want to scope the affordances and limitations of AI-generated explanations, especially around generation techniques, modality, and category. We used the broad definition that an explanation is an answer to a why-question~\cite{lombrozo2011instrumental, lombrozo2012explanation, wilkenfeld2015inference, miller2019explanation}. To \textit{operationalize} the goal, we can adapt some existing guidelines and tailor them for the use-case at hand. For instance, we can use aspects of Liao et al.’s XAI Question Bank~\cite{liao2020questioning} that provides a taxonomy of categories of explanations (global, local, counterfactual, etc.), their respective generation techniques, and connects them to different question types (how, why, what-if, why not, etc.) questions that facilitate user research. Fig.~\ref{fig:framwork} shows a portion of the starter questions that we can use to scope the AI-generated Explanations space.  The \textit{stakeholders} typically tend to be data scientists, XAI engineers (who design the explanations), and end-users (who act on the explanations). 

\subsection{The Social Wing}
This wing of the framework is meant to map out the social requirements of the stakeholders. We partition this wing into \textit{three} interconnected blocks: \textbf{trust in AI} (towards the technology), \textbf{actionability} (at the individual level), and \textbf{values} (at the organizational level). This partition has two motivations: first, it is empirically situated in our case studies. Second, it is also aligned with the three levels of contexts (technical, individual, organizational) made visible by Social Transparency in AI – the perspective that incorporates the socio-organizational context into the conception of AI explainability~\cite{ehsan2021expanding}. This alignment is important because Social Transparency will play a part in connecting the two wings (detailed later) and addressing the sociotechnical gap.

As \citeauthor{ackerman2000intellectual} pointed out, while the technical affordances (what the system can do) is often static, the social requirements are dynamic. The dynamic nature of user needs is primarily why the sociotechnical gap is hard to fully bridge. Often, problems arise on the social side—it could be issued where the AI system is over-trusted (technologically), users have difficulty acting on the AI explanations (individually), there are value tensions and conflicts between stakeholders, etc.

\vspace{0.5em}
\textit{\textbf{Trust (in AI):}}
The goal here is to understand the current level of trust in the AI and scope how we might appropriately calibrate trust. Here, we are looking for situated baseline understandings of user trust in the AI and find points of breakdowns to act on. To \textit{operationalize} the goal, we can use existing surveys around trust in AI like~\cite{hoff2015trust,siau2018building,ashoori2019ai}, interviews guided by the starter question, and Value Sensitive Design (VSD)~\cite{friedman1996value} exercises to get a sense of what aspects of the AI’s performance matters to users, aspects where trust is missing, and how different facets impact the calibration of trust (when to trust the AI’s decision vs. not), etc. Fig.~\ref{fig:framwork} shows some of the starter questions that can guide the investigation. The \textit{stakeholders} of this process are primarily end-users who need to act on the AI’s decision. In the Kuro case, this would be sellers or account managers. While it’s important to engage end-users, collaborating with data scientists and developers (secondary stakeholders) is important because they are often in charge of making changes. 

\vspace{0.5em}
\textit{\textbf{Actionability (of Explanations):}}
This block's goal is to understand what users need to act on the AI's explanation in an informed manner. Actionability is how users act on AI-generated explanations and it plays a role at the decision-making level~\cite{ehsan2021expanding}. Sometimes, even with well-functioning XAI systems, users do not feel empowered to decisive actions. Actionability resides at the junction between trust on technology and organizational values, where conflicted value systems can reduce the actionability of AI explanations. Recall that in the Kuro case, the audits required to override the AI increased the burden. To \textit{operationalize} the goals, we can utilize insights offered by the Social Transparency framework that delineates how social validation (from peers), increased ``peripheral vision'' of past decisions, and support of follow-up actions can boost actionability and AI contestability (Fig.~\ref{fig:framwork} shows some starter questions). To understand how these constructs should manifest in the system design, we can use similar methodological tools in the \textit{Trust} block--- interviews and VSD exercises. The \textit{stakeholders} often include end-users of the system (e.g., analysts in an AI-powered cybersecurity system or mental health specialists in our use case). 

\vspace{0.5em}
\textit{\textbf{Values:}}
The goal here is to understand organizational values and their interplay with individual expectations. Recall that in the Kuro case, the misalignments in what the organization valued and the sellers wanted ultimately led to a breakdown in infrastructure. To \textit{operationalize} the goal, we can utilize methodological tools such as VSD exercises, participatory design workshops, etc. to scope out points of conflict and value alignment between the organization and the individual, understand how accountability is distributed in the Human-AI collaboration, job performance expectations, understand positionality towards AI Ethics, etc. (Fig.~\ref{fig:framwork} has some starter questions). This is a multi-stakeholder process that includes (but is not limited to) managers (in leadership), HR professionals, end-users, and developers. 

While all the building blocks should be visited during the charting process of the gap, there is no prescribed order in which to visit each block. Each block is influenced by others, not just on the same wing but also across the gap. As shared earlier, the sequence of traversal is problem-dependent and insight-driven. Insights from addressing the Actionability block might prompt a visit to the Data block, updating our understanding, and warranting a revisit back to Actionability. 
We can iterate through the blocks, updating our understanding, till we achieve a stable saturation~\cite{saunders2018saturation} of our understanding where further inquiries do not generate novel insights. 
To keep track of our progress and outcomes, we offer three thematic labels (tags)--- \textit{baseline understanding}, \textit{updated understanding}, and \textit{design recommendations}. \textit{Baseline understanding} (BU) occurs the first time we visit a block and address it using the guidelines from our framework to get a sense of the status quo. \textit{Updated understanding} (UU) can occur when revisiting the block with newfound insights from prior processes (e.g., BU of a different block). 
\textit{Design recommendations} (DR) inform the practical development of tools, techniques, and policies (like first-order approximations). DRs can emerge from both baseline and updated understandings.

\subsection{Addressing the Gap: Palliatives \& First-order Approximations}\label{sec:addresssing_framework}

Recall that there is no silver bullet to fully bridge the sociotechnical gap~\cite{munson2013sociotechnical}. Moreover, success does not mean perfection in bridging the gap. 
To address the gap, we can use two approaches what Ackerman calls \textit{palliatives} and \textit{first-order approximations}~\cite{ackerman2000intellectual}.
These approaches are not mutually exclusive and can be used in tandem. 

\textit{Palliatives} are interventions at the political, ideological, and education levels which impact how people perceive the affordances of the system~\cite{ackerman2000intellectual,munson2013sociotechnical}. They are meant to ameliorate the gap without making system-level changes. They can include stakeholder analyses like the value-sensitive design exercises from our case studies. The findings can highlight blind spots in the system, which can then empower people to make informed decisions on usage~\cite{munson2013sociotechnical}. As we saw in Kuro's case study, educational initiatives~\cite{long2020ai,magnussonimproving}) can help bridge the gap where technology fails. As our case studies highlight, a competent system alone is not enough for people to find it useful. Combining analysis to understand where different organizational stakeholders are in the Technology Adoption Life Cycle~\cite{beal1957diffusion,moore1999crossing}, we can use  Technology Acceptance Models~\cite{venkatesh2000theoretical,venkatesh2003user} to understand perceived usefulness of the system. 
These steps can ameliorate certain disconnects between the social and technical elements and inform future technological interventions like first-order approximations. 

\textit{First-order approximations} are tractable solutions that partially solve specific problems with known trade-offs~\cite{ackerman2000intellectual,munson2013sociotechnical}. Success with first-order solutions can involve connecting the social with the technical that \textit{effectively} addresses some of the \textit{most pressing} sociotechnical issues with \textit{known} trade-offs. As our case studies showed, there are \textit{multiple strategies} to develop first-order approximations. 
In Kuro's case, we illustrated how incorporating Social Transparency (ST)~\cite{ehsan2021expanding} acted as a first-order approximation to provide much needed ``peripheral vision'' and promote engagement. By providing visibility of who else did what, when, and why (4W), ST promotes explainability by making context visible at three levels (technological, individual, and organization), which align with the three blocks of the social and technical wings. With conceptual roots in XAI, ST can be a productive first-order approximation to address the gap for XAI systems. Using ST, increased visibility of the sociotechnical factors can provide actionable insights which can potentially address the gap. 
To operationalize ST, we can utilize the constitutive design elements – the 4W (who did what, when, and why) – to encapsulate relevant socio-organizational context.

\subsection{Framework Outcomes} \label{sec:framework_outcomes}
There are two main outcomes or end-products of our framework: (1)~improved understanding of the gap (how the gap looks) and (2)~actionable insights to address the gap (what to do with the gap). The thematic labels above are connected to the outcomes---BU (baseline understanding) and UU (updated understanding) connect to the first outcome while DR (design recommendations) connects to the second one. Our framework provides guidance that facilitates tractable navigation of complex sociotechnical landscapes and produces actionable insights to address the gap. We should have an idea of the most promising places to build a bridge but also where the gap is the largest. Thus, stakeholders are empowered to make informed decisions on what to address in the gap vs. not-- this ability to, which a core tenet of Ackerman’s call to action in addressing the gap~\cite{ackerman2000intellectual}.

%%%%%fr end

%%%%%app start
\section{Applying the Framework}\label{sec:application_framework}
We now share a third case study where we applied this framework in an applied real-world setting. To showcase the versatility of the framework, we use a new domain for the application: \textit{cybersecurity}. The following delineation covers 1.5 years of fieldwork. After the application details, we also share a study investigating the effectiveness of the framework (did the stakeholders find it useful?).

\subsection{Background: Atlas, An AI-powered Cybersecurity System} SecureCorp (pseudonym), a Fortune 500 muti-national cybersecurity technology company, invested \$10 million to build Atlas (pseudonym), an AI-powered cybersecurity system to help analysts manage firewall configurations, especially “bloat” that happens when people forget to close open ports. Over time, as an organization evolves, people forget to close open ports, causing bloat and serious security vulnerability. 
This has high stakes—the wrong port left open could lead to a breach, incurring huge losses for the company (in the order of millions of dollars) because the clients of SecureCorp are primarily banks. On paper, Atlas is impressive---it is purported to have commendable accuracy (92\%); however, the user engagement is only at 2\%. 
Given the significant investment in Atlas and its internal deployment, SecureCorp expected a much higher engagement. Since less than 2\% of the workforce was engaged with Atlas, the question was: why?

\subsection{Problem Identification \& Alignment Using the Framework}
Our framework facilitates a systematic \textit{start} to problem-solving beyond helping us address it. Once we identify the problem, we need to align the problem with the framework: classify the appropriate wing (technical or social) and find the nearest matching building block. Shadowing 10 cybersecurity analysts (hereon, referred to as analysts) for 2 weeks, we identified the problem---\textit{users were not comfortable \textbf{acting on the AI's explanations}}. Checking with our framework, it fell under the \textbf{actionability} block in the \textit{social wing}.

As we visit the conceptual blocks in the rest of Atlas' case study, we use the labels of \textit{baseline understanding} (BU), \textit{updated understanding} (UU), and \textit{design recommendations} (DR) to track the progress of our insights.  

\subsection{Baseline Understanding (BU)}
As noted in Section~\ref{sec:framework_outcomes}, a baseline understanding provides formative information about the problem. It typically occurs during the first visit to a block, where the visits are like fact-finding missions.

\vspace{0.5em}
\noindent{\textbf{Actionability} (Social).} 
To obtain a \textit{baseline understanding}, we conducted interviews with the analysts, guided by the starter questions in our framework. We found that analysts hesitated to act on the explanations because they felt vulnerable and siloed in a vacuum from other colleagues. The nature of work had elements of collaboration that Atlas did not serve. The story of Julie (pseudonym) was infamous: Julie got fired for following the AI’s mistaken recommendation  to close some ports(false-positive). The AI, however, suffered no consequences.

\vspace{0.5em}
\noindent{\textbf{Trust} (Social).}
The vulnerability stemmed from lack of \textbf{trust in the AI} (top of the social wing in Figure~\ref{fig:framwork}). Guided by the framework, we conducted a series of Value-Sensitive Design (VSD) exercises~\cite{friedman1996value,friedman2002value} to get a \textit{baseline understanding} of what is important to them regarding trust. We found two things that informed the next steps: first, accountability was displaced unfairly in the Human-AI relationship. Discussing Julie’s story, another analyst Simon (pseudonym) shared the key concern, “the AI is immune…it has nothing to lose. If the AI is wrong and I trust it, my neck is on the line. So why should I trust something that has no skin in the game?” Second, the analysts could not trust the AI because they lacked an understanding of the technical infrastructure. As Simon said, “[they] really didn’t know what’s going on.” 

These insights prompted a transition to the \textit{technical wing} to get \textit{baseline understandings} of its components. We used the guidelines provided for each block (discussed in~\autoref{sec:details_framework}) to operationally scope the technical affordances and limitations.

\vspace{0.5em}
\noindent{\textbf{Data} (Technical).}
For the \textbf{data} block, we interviewed system administrators and data scientists using our framework around data origins, the purpose of collection, policies, etc. When the development team learned about the analysts’ lack of \textit{trust} and \textit{actionability}, they were motivated to carry out customized exercises following the Datasheets for Dataset framework~\cite{gebru2018datasheets}. We learned actionable information; for instance, that the dataset has proprietary information originally collected for compliance reasons, 
and that there was very little documentation on preprocessing steps, which encouraged the development team to address this deficiency.

\vspace{0.5em}
\noindent{\textbf{Model} (Technical).}
For the \textbf{model}, the data science team shared a substantial limitation---  the model is proprietary and purchased from a vendor (third-party). Thus, SecureCorp's team had no way to ``open'' the ``black-box''. We also learned that this was quite normal---most organizations lack the in-house infrastructure to build systems end-to-end. Often, companies use an ensemble of technologies to provide a service. However, the charting process highlighted blind spots that profoundly impacted \textit{trust} and \textit{actionability}. Upon realizing that they cannot answer our starter questions, the development team had concrete steps to address---their unknowns became known.

\vspace{0.5em}
\noindent{\textbf{Explanation} (Technical).}
Next, we addressed the \textbf{explanation} from Atlas. Using the starter questions, we conducted multiple focus group discussions with the XAI engineers and analysts (end-users). We learned that Atlas generates post hoc explanations (explanation generation after model’s training)~\cite{lipton2018mythos, miller2019explanation, ribeiro2016should, yosinski2015understanding, ehsan2019automated} that are {\em local} (explaining a prediction vs. explaining a model) \cite{guidotti2018local,ribeiro2018anchors,zhang2019interpreting}. The explanations are fine-tuned to the company’s proprietary data. The generation mechanism is model-agnostic to accommodate the lack of access to the vendor’s model. 

While these insights from the social and technical blocks are informative, we cannot act on them if they are not harmonized with the organizational values, which took us back to the \textit{social wing}.

\vspace{0.5em}
\noindent{\textbf{Values} (Social).}
To get a baseline understanding of the interplay between organizational and individual values, we used the framework guidelines to find the points of alignment and conflict through multiple interviews and focus group discussions with leadership, HR representatives, product managers, analysts, and auditors. We found \textit{conflicting values around accountability}; for instance, the organization was set up such that the liability fell on the human’s shoulder, something that hurt actionability (the core problem). Here, we successfully used the insights from previous steps to garner empathy and understanding, mitigating the conflicts between the value systems.

\subsection{Updated Understanding (UU)}
The baseline understanding propelled revisits to certain blocks to update our understanding, which often carried design recommendations. For clarity purposes, we share the updated understanding in this subsection. In the next subsection, we report the design recommendations in one place. We begin with revisits to \textit{data} and \textit{model}.

% \paragraph
\vspace{0.5em}
\noindent{\textbf{Data} (Technical).} 
Conversations with auditors (during the BU of \textit{values}) generated two notable insights applied to the data block. First, we learned that, contrary to our previous understanding, the dataset can be updated yearly (as opposed to every 3 years). Second, the auditors highlighted compliance risks around the lack of pre-processing documentation (problem highlighted during BU of Data). We relayed these findings to the development teams through focus group discussions, which resulted in concrete design recommendations (shared in the next subsection). 

\vspace{0.5em}
\noindent{\textbf{Model} (Technical).}
When the leadership learned (during the BU of \textit{values}) about the lack of access to the model's inner workings, they re-negotiated liability terms with vendors that reduced the burden on analysts, updating our understanding of the model block and generating design recommendations (outlined later).

Using these updated insights, we revisited the \textit{trust} and \textit{actionability} blocks in the \textit{social wing}. 

\vspace{0.5em}
\noindent{\textbf{Trust} (Social).}
Recall that the main reason to get a BU of the technical wing was the \textit{mistrust} in AI due to \textit{lack of understanding} of the technical blocks. We relayed the  BU and UU of data, model, and explanation to the analysts. The increased visibility of what was possible and not possible alone recalibrated their \textbf{trust} in AI. For instance, learning about the new data update policies calibrated their expectations on the correctness of AI decisions on the latest security threats. The re-negotiated model liability with the vendor provided psychological relief.

\vspace{0.5em}
\noindent{\textbf{Actionability} (Social).} 
The recalibration of trust updated \textbf{actionability} given the close relationship between them.  Recall that analysts felt vulnerable and siloed. We wanted to address this problem. Guided by the framework, we utilized the lenses of Social Transparency (ST) to increase the ``peripheral vision'' of decision-making by providing the context of past decisions. Interviews revealed that analysts strongly resonated with the idea of reducing individual vulnerability by distributing accountability through increased visibility of each other’s interactions. We used the guidelines from ST and engaged in participatory design sessions to operationalize \textit{how} the 4W (who did what, when, and why) of ST could manifest at a design level to promote informed actionability. Aligned with the ST guidelines, we learned how the increased socio-organizational context can promote peer-to-peer learning, which adds social validation to individual decision-making. These led to design recommendations (detailed in the next subsection). 

After these sessions, we reached a saturation~\cite{saunders2018saturation} of our understanding where further investigations did not generate novel information. We proceeded to make design recommendations.

\subsection{Design Recommendations (DR)}

The updated understanding produced design recommendations at technological and policy levels. 

On the \textit{\textbf{technological}} side, there are recommendations on \textbf{explanation} and \textbf{data}. On the \textbf{explanation} side, analysts wanted the 4W design insights (derived from the UU of Actionability) to be incorporated alongside the AI's explanation. Next to each AI-generated explanation, they wanted the socio-organizational context---who else did what with similar AI decisions in the past and why (the 4W). There were concrete design manifestations of the 4W: for \textit{who}, analysts wanted years of experience to be displayed to situate the authority behind their recommendation. They voted against revealing names to protect privacy. For \textit{what}, they wanted not just what the past decision was (accept/reject) but also what the outcome was (threat nullified/not). For \textit{when}, beyond the date, they wanted more temporally relevant information such as last dataset update. For \textit{why}, they collaboratively agreed on a standardized template for consistency purposes. On the \textbf{Data} side, given the new understanding of update policies and pre-processing details, this information should be tagged along with other details alongside the \textit{when} in the 4W. 

In terms of \textit{\textbf{policies}}, during the BU of \textbf{value}, we productively engaged with leadership. Empowered by our findings, they launched an educational campaign (a palliative measure~\cite{ackerman2000intellectual,munson2013sociotechnical}) led by the analysts (the end-users) that reformed policies and training procedures. On the \textbf{model} end, we revised employee training programs to cover the re-negotiated liability contract with the vendor so that new analysts do not suffer Julie's fate. These policy changes align with one of Ackerman’s proposed methods---educational interventions---to address the sociotechnical gap~\cite{ackerman2000intellectual}.

\subsection{Framework Outcomes}
Harmonizing user requirements, organizational incentives, and technical affordances, we worked with the development team to implement the design recommendations (4W of Social Transparency and policy changes). Within one month of deployment, \textit{the engagement was 93\%, a drastic change from the initial 2\%}. We continued to track progress over the next four months through interviews, surveys, and workshops. The results were promising---the improved engagement (94\% on average) addressed the actionability problem. Similar to the Kuro case study, the increased ``peripheral vision'' afforded by the 4W of ST (who did, what, when, and why)~\cite{ehsan2021expanding} helped
analysts gain a deeper understanding of the socio-organizational context, which appropriately calibrated their trust on the AI and self-confidence to act on the AI's recommendation. The increased visibility of the technical affordances enabled informed opinions about the limitations of Atlas. Most importantly, analysts reported an improvement of Atlas’ explainability because it was no longer devoid of the socio-organizational context that was necessary for informed actionability.

\subsection{\edit{Investigating Framework Effectiveness}} \label{sec:framework_effectiveness}

\edit{While the engagement gains are important for SecureCorp, as a research team, we wanted to investigate the following:  how effective was the framework? Did the stakeholders find the framework useful? To answer these questions, we conducted a follow-up study with the following procedure.}

\subsubsection{\edit{Methods}}
\edit{We hosted a two-day workshop attended by 23 participants (referred to as P1-P23 below) with diverse backgrounds (XAI engineers, Product Managers, UX Researchers, HR managers, Responsible AI managers, etc.). The workshop lasted 5 hours each day. We analyzed 10 hours of recorded footage through the lenses of thematic analysis~\cite{braun2006using}. Using an open coding scheme, we produced in-vivo codes (directly from the data) and clustered them themes. We iterated till we reached stability in our codes and respective themes, grouping them at a topic level using a combination of mind-mapping and affinity diagramming.}

\subsubsection{\edit{Findings}} \edit{Below we present our findings on the effectiveness of the framework, related challenges, and how we may address the challenges.}

\edit{\textbf{The framework empowered participants to appreciate the multi-stakeholder and sociotechnical nature of the problem, expanding their understanding of explainability.}
Participants highlighted how the framework helped them appreciate that “everything is situated, nothing happens in a silo, and we all need to depend on each other” (P3). Reflecting on a newfound sociotechnical perspective, participants with engineering backgrounds appreciated that “it’s a multiplayer game, and [they] were playing it a 1-player game” (P2). They also expressed welcomed relief at ``not being alone to handle all the complex issues of AI like fairness, bias, explainability'' (P23). “The framework helped [them] see that it is unfair to expect engineers to carry the entire weight of AI explainability because the problem is not just technical, it’s sociotechnical” (P14). Most importantly, a sociotechnical lens afforded participants to view “XAI as something more than model transparency” (P8). This participant captures the newfound realization: 
\begin{quote}
\small
    ``It’s sad that I had such a very narrow view of XAI where XAI was equal to algorithmic transparency. AI systems are made of people and algorithms. I now realize explainability \textit{of the system} is much more than explainability \textit{of the model}...
    This is why [engineers] like me need to know where the data comes from, what the model cannot do…otherwise, how can I help my end user to know \textit{when} to rely on the AI’s explanation vs. not?'' (P12, XAI engineer, emphasis added)
\end{quote}}

\edit{\textbf{Participants emerged with a deeper understanding of the sociotechnical gap in XAI.}
Before participants went through the framework, they “lacked a clear understanding of the gap and how the technical and social sides mattered” (P10). The explicit sociotechnical focus with “blocks dedicated to each side made the process modular and traceable” (P14). When asked how the framework augmented their understanding of the gap, this participant summarized it well:
\begin{quote}
\small
    ``The process provided a high-resolution image of the gap as opposed to a very blurry picture when we started. We know where the gap is the widest and the closest. This knowledge is crucial for interventions. Even as someone with a PhD and who knew Ackerman’s work, it never really dawned on me that we could address XAI issues like this. '' (P15, UX Researcher)
\end{quote}}

\begin{table}[t]
\sffamily
\caption{\edit{Framework effectiveness workshop participant details}}
\vspace{-3pt}
\label{table:participant details}
  \begin{minipage}[t]{0.5\columnwidth}
  \centering
    \footnotesize
	\setlength{\tabcolsep}{5pt}
    \begin{tabular}{clc}
 \textbf{ID}&\textbf{Organizational Role}&\textbf{Experience} \\
 \toprule
 \rowcollight \multicolumn{3}{c}{\textit{Engineering and Data Science}}\\
P1   & XAI Engineer   & \textgreater 5 ys.\\
P7   & XAI Engineer     & \textgreater 10 ys.\\
P12  & XAI Engineer                        & \textgreater 4 ys.\\
P23  & XAI Engineer                        & \textgreater 3 ys.\\
\hdashline
P2   & Data Scientist  & \textgreater 10 ys.\\
P6   & Data Scientist   & \textgreater 5 ys.\\
P14  & Data Scientist                      & \textgreater 6 ys.\\
 \rowcollight \multicolumn{3}{c}{\textit{User Research}}\\
P3   & UX Researcher   & \textgreater 5 ys.\\
P4   & UX Researcher   & \textgreater 5 ys.\\
P15  & UX Researcher                       & \textgreater 10 ys.\\
\hdashline
P16  & UX Designer                         & \textgreater 13 ys.\\
& & \\
% \hdashline
% \hdashline
\bottomrule
\end{tabular}
  \end{minipage}\hfill
    \begin{minipage}[t]{0.49\columnwidth}
  \centering
    \footnotesize
	\setlength{\tabcolsep}{4pt}
    \begin{tabular}{clc}
 \textbf{ID}&\textbf{Organizational Role}&\textbf{Experience} \\
 \toprule
 \rowcollight\multicolumn{3}{c}{\textit{Product Management}}\\
P5   & Product Manager  & \textgreater 15 ys.\\
P8   & Product Manager   & \textgreater 8 ys.\\
P9   & Product Manager  & \textgreater 11 ys.\\
 \rowcollight\multicolumn{3}{c}{\textit{Organizational Governance}}\\
P10  & Responsible AI Manager/ AI Ethics & \textgreater 3 ys.\\
P11  & Responsible AI Manager/ AI Ethics & \textgreater 4 ys.\\
P19  & Responsible AI Manager/ AI Ethics & \textgreater 3 ys.\\
P20  & Responsible AI Manager/ AI Ethics & \textgreater 5 ys.\\
\hdashline
P13  & Compliance Officer                  & \textgreater 10 ys.\\
P22  & Compliance Officer                  & \textgreater 11 ys.\\
\hdashline
P17  & VP of Technology                    & \textgreater 15 ys.\\
P21  & SVP of Technology                   & \textgreater 20 ys.\\
\hdashline
P18  & HR Manager                          & \textgreater 16 ys.\\
% \hdashline
% \hdashline
% \hdashline
% & & \\
\bottomrule
\end{tabular}
  \end{minipage}
\Description[Table showing participant details]{In this table, we share the participant details. From left to right, the columns are Participant ID, Role, Domain, and Years of Experience}
\vspace{-1em}
\end{table}

\edit{\textbf{Blind spots in existing organizational practices were revealed. }
Participants expressed that the “framework allowed [them] to clearly highlight blind spots” (P19). For example, while obtaining the BU of the Values block we found that there were value tensions around how accountability was structured and how analysts felt vulnerable, which impacted their actionability, and hurt engagement (the root problem). Most importantly, “the framework helped [stakeholders] see that [they] needed the 4W—the social transparency—and that technical transparency wasn’t cutting it” (P6). Participants felt that “had [they] not gone through the process, these blind spots may not have been identified and addressed” (P21).}

\edit{\textbf{The collaborative process led to improvements in inter-departmental dynamics.}
One of the “best effects of going through the blocks has been improvement in [participants’] relationships with other departments. In large organizations like SecureCorp, work is often siloed along verticals. Participants shared that the framework “provided peripheral vision of who else was there, how they can help, and most importantly, brought everyone under the same roof to solve the joint-problem” (P17). Even though “the start was a rough one and [non-engineering participants] didn’t see eye to eye with the engineering team, [they] now have a better working relationship and mutual understanding” (P15). The process, participants expressed, “brought people from different bubbles together, which is an amazing achievement in a complex organization” (P22).} 

\edit{\textbf{The process activated latent, pre-existing organizational information into actionable knowledge.}
Being able to re-use and re-purpose existing practices and guidelines was a major win for us in terms of getting buy-in and adoption. As shared in~\autoref{section:casestudies}, we made a conscious effort to ensure we leverage as much of the existing infrastructure as possible; for instance, we leveraged familiar HCI techniques and AI guidelines like Model Cards that already had built-in adoption. Product Managers particularly appreciated this “sustainable way of thinking because it made the most out of what [the company] already had” (P9). These choices minimized extra burden on our participants, which they appreciated. It also activated latent information into actionable knowledge without adding significant burden to stakeholders, elaborated by this participant:
\begin{quote}
\small
    ``Thanks for not forcing us to do yet another new checklist. We’re often jaded of researchers telling us to do new things without looking at how we can leverage what we've! I'm really happy that we got a lot of mileage from existing information and processes. I loved how this process helped us see old things in a new light and solve a really sticky problem.'' (P17, VP of Technology)
\end{quote}}

\edit{\textbf{Beyond benefits of using the framework, participants highlighted two main challenges. }
The \textbf{first challenge} is around \textbf{``the issue of ownership and responsibility of each block''} (P1). Participants shared that they were expecting more guidance from the framework. Those from the Responsible AI and Policy teams, wondered: ``Since it’s a multi-stakeholder problem, it’s not clear who owns which block. Who is ultimately accountable? Without clear guidance, we may end up playing hot potato or resort to `not my problem'-type thinking'' (20). 
\textit{When asked how we may alleviate that hurdle}, we had generative and critically constructive suggestions. First, given problem contexts can vary, participants agreed it was ``impossible to decide [ownership] before-hand, so it should not be top-down'' (P22). Issues of responsibility should be discussed “when [stakeholders] have a clear idea of the problem'' (P4). Second, since each block in the framework often has multiple stakeholders, ownership “need not be on only one team; it can be shared'' (P12). The relative split of accountability for the block can be decided by asking---``which group has the most to lose if things go wrong in this block?''. Leadership ``may need to proactively engage in cases of disagreements'' (P20). Last, the conversations and should start early---``before embarking on a baseline understanding and revisited during updated understanding to see if things have changed'' (P2). }

\edit{\textbf{Another challenge} is around the \textbf{complications of obtaining collective buy-in during the early stages} of the framework. 
Here, teams struggled to get buy-in during the early stages and wanted the framework to better steer the process. Data scientists and developers ``initially struggled to understand why non-developers had to be involved when discussing the model'' (P7)
\textit{Participants offered constructive solutions to this problem of buy-in. }They suggested to add ``an all-hands meeting with team leads during onboarding. To get buy-in, it’s important that value propositions are clearly laid out—what’s in it for each team and what we do lose if we don’t put our heads together?'' (P18). To give stakeholders a shared sense of ownership to facilitate the buy-in, participants advocated and appreciated our participatory methods. They “felt included; it was clear that [they] were problem-solving \textit{with} [the researchers]'' (P16). This Product Manager put it succinctly: ``if you want people to join you, first you need to make them feel like their voices matter; next you make it crystal clear early on why it’s a team sport and we all need to pull our weight”(P9)}

%%%%%app end

%%%%%disc start
\section{Discussion and Implications} \label{sec:implications}

\subsection{Transferring the Framework to New Domains}

We constructed the framework with the ethos of customizability and versatility, striving to balance specificity with generality. This is why we developed it from distinct domains and applied it in a third new domain. The diversity of domains is an asset and speaks to the versatility of the framework’s applicability. Despite their diversity, they share a vital common element: \textit{all domains have the sociotechnical gap}. Thus, the framework can apply to domains beyond sales and cybersecurity like robotics, autonomous driving, etc. As long as the problem has an explainability component embedded in sociotechnical environments, the framework can be used to chart the sociotechnical gap. This is because the answers to the questions we ask and how we operationalize each block, and the starter questions are similar. 
Moreover, when transferring the framework to new domains, we should be mindful of whether the contexts are permitting. 
Inherent in transferability is context-sensitivity~\cite{hayes2011relationship,stringer2007action,finfgeld2010generalizability}. Real-world XAI settings have different contexts, domains, and application area---as such, the transferability of our framework to new domains is subject to context-sensitivity. Given its construction and the existence of the sociotechnical gap in all sociotechnical systems, our framework is likely to have broad transferability but not blanket generalizability. 

We offer \textit{three} “knobs” (entities) to consider as we calibrate the framework to new domains: \textit{nature of collaboration, infrastructural complexity,} and \textit{stakes (or consequences)}. These three knobs are not the only ones that can potentially inform transfer and are not meant to be exhaustive. They are, however, informative and provide guidance during transfer.

For the \textit{nature of collaboration}, the higher the collaborative or cooperative nature of work, the higher the impact from the social wing of the framework, and the higher the potential impact of Social Transparency in bridging the social and technical factors. The level of collaboration in our use case is relatively high---in all three cases, we had multiple users interacting with the AI system. In contrast, if the use case is geared towards 1-1 Human-AI interaction, then the social factors might not weigh as much. Beyond the number of users involved in the Human-AI interaction, we should also consider the nature of geographic co-location and organizational culture. A more distributed workforce (e.g., multi-nationals) might require higher levels of Social Transparency. 

For the \textit{infrastructural complexity}, we should consider complexities both at the technical and social (organizational) levels. Technical complexities can include complex data infrastructures, models (ensembles of deep neural networks), etc. On the social side, complex organizational structures can impact how we understand the interplay of values, how actionability works, and what type of human factors govern trust in AI. The level of complexity can correlate with the number of “passes” or iterations we need to do over each block---the higher the level of complexity, the more time/effort we might need to invest in operationalizing each block (e.g., multiple Model Cards reporting for ensemble models), and the higher the chances of multiple iterations for updated understandings. This is why  our framework provides thematic labels (baseline understanding, updated understanding, etc.) to manage progress and track challenges.

For the \textit{stakes} (or consequences), we should be mindful of the eventual impact on human lives. These can range from distributive justice issues like the allocation of relief funds (algorithmically) to users risking job losses due to accepting a false positive AI decision. An AI system responsible for recommending music has a different level of stakes than one that recommends loan approvals. For example, in the Atlas (cybersecurity) case study, the stakes were high---a missed step could cost someone’s job (case in point, Julie’s story). The liability of the AI’s error resided entirely on the human. This displacement of accountability created a sensitive situation than required careful stakeholder management. Not only did we have to revisit blocks (e.g., \textit{actionability}) in the social wing to update our understanding, we also had to be careful in operationalizing the \textit{values} block---when engaging with leadership, we had to strike a balance where no single party felt blamed for the situation. Thus, there are socio-political aspects to consider when considering stakes. 

\vspace{-1em}

\subsection{Practical Implications}
Our work bears practical implications that can help researchers and practitioners (stakeholders on the technology development end such as data scientists, AI engineers, technical product managers, UX Researchers) operationalize our proposed framework. 

\subsubsection{Implications for Practitioners}
\textit{First,} thinking sociotechnically pays dividends— technology problems in organizational environments are sociotechnical. This entails that they have both social and technical components. Thus, technical interventions alone cannot solve sociotechnical problems. When looking for solutions, developers and data scientists are likely to make progress if they consider both social and technical dimensions.
\textit{Second,} the problem space requires multi-stakeholder engagements; thus, practitioners should actively seek collaboration with other teams. As we highlighted in the framework effectiveness study (\autoref{sec:framework_effectiveness}), engineers and developers were relieved to not feel alone in handling sociotechnical complex issues in AI. The framework helped them realize that ``it's a multi-player game'' (P2) and that a purely technical perspective imposes an unfair burden on engineers to solve problems that are multi-stakeholder and sociotechnical. 
\textit{Third,} there is value in going through the activity of charting the gap even though the sociotechnical gap may never be fully bridged. As we saw in \autoref{sec:application_framework}, from highlighting organizational value tensions to unearthing non-obvious technical limitations, the framework-driven process provided actionable insights on not just understanding the sociotechnical gap in XAI but also how to address it. The process also had the positive side-effect of surfacing organizational barriers that may hinder cross-functional teamwork that, if addressed formally or informally, can improve future best practices.

\vspace{-1em}
\edit{
\subsubsection{Implications for Researchers}
\textit{First,} taking a participatory approach helps get stakeholder buy-in---if stakeholders feel that they are designing \textit{with} the researchers, rather than the researchers designing \textit{for} them, they tend to feel included and take engaged ownership of the process. We framed all problems as cooperative joint problems~\cite{aarikka2012value} where the power dynamics were calibrated as much as feasible. Being participatory can also complicate collective buy-ins (a challenge identified in our effectiveness study); e.g., who calls the shots for a given block. We share mitigation strategies in~\autoref{sec:application_framework} such as clearly articulating value propositions before starting the process. 
\textit{Second,} researchers should leverage existing information and documentation before introducing new checklists and processes. As our framework effectiveness workshops showcased (in~\autoref{sec:framework_effectiveness}), such a sustainable approach improved (a)~the chances of stakeholder buy-in and acceptance, (b)~activated latent information into actionable knowledge without adding a significant burden to stakeholders, and (c)~mitigated what participants called \textit{checklist burnout} where people feel overwhelmed with long checklists and lose engagement. 
This is why our framework encourages using established checklist guidelines and familiar experimental methods.
\textit{Third}, where feasible, we suggest engaging in group settings that have lower barriers to participation because they tend to be generative. Relatively low-stakes group settings like ``lunch-n-learns" where staff gather to listen to presentations are resourceful avenues to get initial interest in a project. Given the relevant stakeholder(s) might already be there, it reduces the barrier to getting people in a room. This is why, whenever possible, we conducted workshops. We often recruited through lunch-n-learn sessions and used virtual workshops to reduce the barrier to participation.
}
\edit{
We hope that these practical implications will help future researchers and practitioners operationalize the framework in real-world settings.
}

\subsection{Conceptual \& Theoretical Implications}
By situating the sociotechnical gap in the context of XAI, our framework adds to its epistemology by \textit{expanding the conceptual landscape} in three ways. \textit{First,} it showcases the \textit{utility of conceptually understanding the gap} by connecting it with practical outcomes. By providing a systematic process-driven structure to chart the gap, our framework showcases how the very act of understanding the gap can produce design interventions to address the gap. This connection motivates the practical utility in the pursuit of~\citeauthor{ackerman2000intellectual}’s main goal in proposing the gap—the need to understand it~\cite{ackerman2000intellectual}. It also speaks to the shift in focus we advocated in the introduction---one that goes from a “gap filling” philosophy to a “gap understanding” one when addressing the sociotechnical gap in XAI.

\textit{Second,} our framework \textit{connects diverse threads} of work in AI and HCI and \textit{translates} them in the \textit{context of XAI}, adding to its conceptual landscape. For example, on the technical wing, we integrate existing guidelines (like Datasheets for Datasets~\cite{gebru2018datasheets}, Model Cards~\cite{mitchell2019model}, XAI Question Bank~\cite{liao2020questioning}, etc.) that previously existed in their own niches. On the social wing, we derive translational insights from XAI paradigms (like Social Transparency~\cite{ehsan2021expanding}) that incorporates socio-organizational elements. More importantly, the framework delineates the interplay between the building blocks of each wing as well as offers methodological guidelines (like Value Sensitive Design (VSD)~\cite{friedman2002value} and Participatory Design (PD)~\cite{schuler1993participatory}) on operationalizing the goals. 

\textit{Third,} by drawing boundaries that include the social factors rather than purely technical ones, our framework reframes the conceptualization of the relevant factors, potentially empowering stakeholders to \textit{avoid certain techno-centric traps} such as Solutionism ~\cite{morozov2013save} and Formalism~\cite{green2020algorithmic}. Selbst et al.\cite{selbst2019fairness} highlight falling into these traps (by excluding the social context) can cause AI interventions to exacerbate societal inequities and unfairness. 
By ``refocusing of design in terms of process rather than solutions''~\cite{selbst2019fairness}, our framework can help researchers and practitioners avoid these pitfalls. It encourages a sociotechnically-informed ``heterogeneous engineering approach''~\cite{selbst2019fairness}, one that critically reflects on \textit{what} social factors to use \textit{why}, and \textit{when} to implement design interventions ~\cite{green2020algorithmic,selbst2019fairness}. For example, in the Atlas case study, the framework not only helped us devise design recommendations around the 4W (who, what, when, why) of Social Transparency~\cite{ehsan2021expanding}, but it also highlighted the areas (actionability and trust) in which it is most impactful. 

The conceptual implications are connected to our \textit{design philosophy}. As we shared in the introduction, our framework is generative, not normative. Its process-driven nature is meant to empower stakeholders (e.g., researchers, designers, developers) to systematically chart the sociotechnical gap and generate insights to address the gap. We do not anticipate there exists a one-size-fits-all paradigm that could chart the gap universally for all contexts. There could be many paths that lead to the same destination. As such, we do not impose restrictions on the exact path to mapping the gap. Instead, we focus on operationalizing the building blocks systematically where the generative insights guide the path-finding process. For example, a startup company might take a different sequence of actions in charting the gap compared to a large multinational organization. While they might have different types of gaps, the gap still exists, which is where our generative framework can help in both cases. Incorporating the lenses of Social Construction of Technology developed by \citeauthor{pinch1984social}~\cite{pinch1984social}, we
encourage \textit{interpretive flexibility} of the process from different \textit{relevant social groups}, each with its own goals and visions of the technology. This design philosophy not only permeates this framework but also has conceptual implications on future work based on it.

 \subsection{Limitations \& Future Work}
We have taken a formative step on how to chart the gap and address it by deriving an analytical framework using two case studies and applying it on a new one. Given this first step, the insights from our work should be scoped accordingly. 
We need to do future work on applying the framework across more domains and in different settings.
In our case studies, both the development and end-users were part of the same organization. 
Future work could explore contexts where the development and end-user teams are from different organizations (e.g., service providers and customers). 

For future iterations, we are inspired by Agre's design philosophy of Critical Technical Practice~\cite{agre1997toward}.
Aligned with Agre’s notion, “at least for the foreseeable future, [we] will require a split identity – one foot planted in the craft work of design and the other foot planted in the reflexive work of critique.”~\cite{agre1997computation}. As such, our exploration of the sociotechnical gap in XAI, at least for the foreseeable future, is a work-in-progress.
We have “planted one foot" in the work of design by offering a framework that helps map the gap and address it. Now, we seek to learn from and with the broader HCI and XAI communities as we “plant the other foot” in the self-reflective realm of critique.

%%%%%disc end

%%%%%conc start
\section{Conclusion}

Explainable AI (XAI) systems are inherently sociotechnical; thus, they are subject to the sociotechnical gap. This paper targeted the problem of charting the sociotechnical gap between the technical affordances and the social needs in XAI systems. We used two case studies in two different domains (sales and mental health) to empirically derive an analytic framework to facilitate charting the sociotechnical gap in XAI. This framework consisted of three building blocks in each of the technical and social wings. The technical wing included data, model, and explanations components, and the social wing included trust, actionability, and values components. For each of these building blocks, we provided a set of starter questions that would help address the sociotechnical gap in XAI systems. Further, we delineated how to address the gap through the lenses of palliatives and first-order approximations~\cite{ackerman2000intellectual}. Next, we applied the proposed framework to a third case study in a new domain (cybersecurity) to showcase its affordances, and how it helped to not only understand the problem but also to generate actionable insights to adequately address the problem and improve the system's explainability. Our work provided design recommendations and potential challenges, risks, and tensions in mapping out and addressing the sociotechnical gap in XAI.

\begin{acks}
With our deepest gratitude, we acknowledge the time our participants generously invested in this project. Without their input, this project would not have been possible. We also want to thank the organizations, the sites for the case studies, for their cooperation. We are grateful to members of the Human-Centered AI Lab and The Social Dynamics and Well-Being Lab at Georgia Tech whose continued input refined the conceptualizations presented here. 
We are indebted to Vera Liao, Michael Muller, and Samir Passi for their generous feedback that helped scope the project appropriately.
Special thanks to Rachel Ehsan for generously providing proofreading feedback. 
This project was partially supported by the National Science Foundation under Grant No. 1928586.
\end{acks}

%%%%%conc end

% \input{2relatedwork}
% \input{3casestudies}
% % \input{3casestudies2}
% % \input{4reinterpretation}
% \input{4framework}
% \input{5application}
% \input{6discussion}
% \input{7conclusion}

\bibliographystyle{ACM-Reference-Format}
\bibliography{references,MDC_Refs1,MDC_Refs2,sample-base,STPaperRefs,refs2}

\end{document}
\endinput
%%
%% End of file `sample-authordraft.tex'.